\documentclass{aa}  

\usepackage{graphicx}
\usepackage{txfonts}
\usepackage{gensymb}
\usepackage{xcolor}
\usepackage{appendix}
\usepackage{placeins}
\usepackage{natbib,twoopt}
\usepackage[breaklinks=true]{hyperref} 
\bibpunct{(}{)}{;}{a}{}{,} 
\makeatletter
\newcommandtwoopt{\citeads}[3][][]{\href{http://adsabs.harvard.edu/abs/#3}%
{\def\hyper@linkstart##1##2{}%
\let\hyper@linkend\@empty\citealp[#1][#2]{#3}}}
\newcommandtwoopt{\citepads}[3][][]{\href{http://adsabs.harvard.edu/abs/#3}%
{\def\hyper@linkstart##1##2{}%
\let\hyper@linkend\@empty\citep[#1][#2]{#3}}}
\newcommandtwoopt{\citetads}[3][][]{\href{http://adsabs.harvard.edu/abs/#3}%
{\def\hyper@linkstart##1##2{}%
\let\hyper@linkend\@empty\citet[#1][#2]{#3}}}
\newcommandtwoopt{\citeyearads}[3][][]%
{\href{http://adsabs.harvard.edu/abs/#3}
{\def\hyper@linkstart##1##2{}%
\let\hyper@linkend\@empty\citeyear[#1][#2]{#3}}}
\makeatother

\def\gaia{\textit{Gaia}\xspace}
\def\gdr3{\textit{Gaia}~DR3\xspace}

\def\bp{$G_{\rm BP}$\xspace}
\def\rp{$G_{\rm RP}$\xspace}

\begin{document}

\titlerunning{\gaia~method paper: analysis of flares} 

\title{\gaia method paper: an algorithm to detect flare events in \gaia time-series. }

   \author {E. Distefano
          \inst{1}
          \and
          A. C. Lanzafame
          \inst{1,2}
          \and
          A. F. Lanza
          \inst{1}
          \and S. Messina
          \inst{1}
          \and I. Pagano
          \inst{1}
          \and M. Audard
          \inst{3}
          \and G. Jevardat de Fombelle
          \inst{3}
          \and
          B. Holl
          \inst{3}
          \and I. Lecoeur-Taibi 
          \inst{3}
          \and N. Mowlavi
          \inst{3}
          \and K. Nienartowicz
          \inst{3,5}
          \and L. Rimoldini
          \inst{3}
          \and D. W. Evans,
          \inst{4}
           \and M. Riello
        \inst{4}
        \and P. García-Lario
            \inst{6}
          \and P. Gavras
          \inst{7}
        \and L. Eyer
          \inst{3}
           }

   \institute{ INAF - Osservatorio Astrofisico di Catania\\
              Via S. Sofia, 78, 95123, Catania, Italy\\
              \email{elisa.distefano@inaf.it}
            \and University of Catania, Astrophysics Section, Dept. of Physics and Astronomy\\
          Via S. Sofia, 78, 95123, Catania, Italy
          \and Department of Astronomy, University of Geneva, \\ Chemin Pegasi 51, 1290 Versoix, Switzerland
          \and Institute of Astronomy, University of Cambridge,\\ Madingley Road, Cambridge CB3 0HA, United Kingdom
          \and Sednai Sàrl, Geneva, Switzerland
           \and European Space Agency (ESA),European Space Astronomy Centre (ESAC), Camino bajo del Castillo,\\ s/n,Urbanizacion Villafranca del Castillo, Villanueva de la Ca\~nada, 28692 Madrid, Spain
          \and Starion for European Space Agency (ESA), Camino bajo del Castillo,\\ s/n,Urbanizacion Villafranca del Castillo, Villanueva de la Ca\~nada, 28692 Madrid, Spain
       }

   \date{Received ; accepted }

 
\abstract
{
{The third \gaia\ data release (GDR3) includes  the {\tt gdr3\_rotmod} catalogue of 474\,026 stars with variability that is attributed to magnetic activity analogous to that observed in the Sun, arising from rotational modulation induced by spots and faculae. These stars are therefore expected to exhibit flaring activity driven by magnetic reconnection processes.}
{We aim to demonstrate that stellar flares can be reliably identified and characterised in the sparsely sampled \gaia\ photometric time series.}
{We developed and validated a dedicated flare-detection pipeline that exploits the simultaneous multi-band photometry in the $G$, $G_{\rm BP}$, and $G_{\rm RP}$ bands. The algorithm identifies outliers associated with a concurrent increase in brightness and blueing of the stellar colour, and applies consistency criteria to distinguish genuine flare events from instrumental or calibration artefacts.}
{Applying this method to the {\tt gdr3\_rotmod} catalogue, we detect 3\,217 flares occurring on 2\,818 stars out of 474\,026 analysed sources. All events are provided in a dedicated catalogue together with their photometric amplitudes. For a subset of 348 flares, we estimate an effective temperature using a black-body approximation, and we identify 29 hyper-flares with amplitudes $A(G)\ge0.75$\,mag, which occur preferentially in M dwarfs.}
{Despite the sparse temporal sampling of \gaia, our results demonstrate that its multi-band photometry and spectrophotometric information enable the robust detection of flares and the basic characterisation of their properties across an all-sky stellar sample. This work establishes the methodological foundation for future \gaia\ flare catalogues and provides a homogeneous, complementary view of stellar flaring activity alongside high-cadence missions such as {\it Kepler} and TESS.}
}


   \maketitle
%

\section{Introduction}
Flare events are variability phenomena occurring in magnetically active stars. A typical flare is  characterised by an impulsive enhancement of the stellar flux followed by a slow decay phase  with time-scales ranging from few minutes to several hours. Such an event is caused by  magnetic reconnection i.e. a re-arrangement of the topology of the stellar magnetic field  \citep[see e.g.][]{1989SoPh..121..299P,2010ARA&A..48..241B} during which the magnetic energy stored in the field is converted into plasma heating, acceleration of charged particles, and radiative emission. 
Flares showing an enhancement of the continuum emission  in the visible spectrum are called white-light flares (WLFs). 

Space-based photometric missions such as {\it CoRoT} \citep[Convection, Rotation and planetary Transits][]{2006cosp...36.3749B}, {\it Kepler} \citep{2010Sci...327..977B}, K2 \citep{2014PASP..126..398H} and TESS \citep[Transiting Exoplanet Survey Satellite][]{2015JATIS...1a4003R}  have enabled the detection of WLFs in thousands of stars and the investigation of their statistical properties.
The analysis of data coming from these surveys allowed us to see how the energy released during flare events, the duration, and the frequency at which they occur are related to the main stellar parameters. These studies indicated, for example, that M-type stars flare more frequently and with a stronger enhancement with respect to the quiet flux than K-type stars. Also, flares occurring in M-type stars have on average a shorter duration than  in K-type stars \citep{2011AJ....141...50W}.
These statistical works provide constraints on the topology of stellar magnetic fields and on dynamo models.
 .

The study of stellar flares is also becoming more and more important in the field of the exoplanets science. Indeed the energy and the occurrence rate of flares in a given star are crucial parameters to establish the habitability of the rocky planets hosted by the star \citep{2006Icar..183..491B}. In fact, the UV radiation emitted during flare events can trigger the reactions needed to synthesise ribonucleotides that are molecules essential to start the prebiotic chemistry \citep{2015EPSC...10....1S}. On the other hand, an excessive amount of UV radiation   can remove the ozone layer surrounding the planet making its surface inhabitable \citep[see e.g.][]{2019AsBio..19...64T}. 

The amount of UV radiation released during a flare is usually estimated by modelling the flare continuum emission with a black-body radiation corresponding to a temperature between $T=9\,000 K$ and $10\,000 K$ \citep[see, for instance,][]{2020AJ....159...60G}. This first order approximation is based on multi-band or spectroscopic observations of tens of stars \citep[see e.g.][]{2003ApJ...597..535H,2013ApJS..207...15K}.
However, this assumption has been questioned by recent studies. Indeed,  \cite{2020ApJ...902..115H} analysed the emission of 44 flares in  27 K5-M5 dwarfs and showed that the temperatures of these flares range between 6\,600 and 43\,000 K. Recently, \cite{2022AJ....164..223R} employed the multi-band measurements of {\it CoRoT} to estimate the black-body temperature of 209 WLFs in 69 F-K dwarfs and found that the average temperature of these flares is 6\,400 K with a standard deviation of 2\,800 K.

We investigate the capability of the \gaia\ photometric time series in the $G$, \bp, and \rp\ bands to detect white-light flares and constrain their properties. Although \gaia\ was not designed for high-cadence monitoring and provides sparse temporal sampling, its multi-band and multi-epoch photometry enables the identification and characterisation of flare events.

Section~2 describes the data analysed in this paper. Section~3 presents the method adopted to detect and characterise flares. The results are presented and discussed in Section~4, and the main conclusions are summarised in Section~5.

\section{Data}

\subsection{Input sample}

The targets analysed in this work are selected from the \gaia\ DR3 table 
{\tt gaiadr3.vari\_rotation\_modulation} (hereafter {\tt gdr3\_rotmod}). 
This table contains 474\,026 variable sources for which the \gaia\ {\tt rot\_mod} 
pipeline \citep{2023A&A...674A..20D} detected rotational modulation, that is, 
flux variability arising from the combined effect of stellar rotation and the 
non-uniform distribution of spots and faculae over the stellar photosphere.

The sample primarily consists of main-sequence stars later than spectral type F5, 
and also includes RS CVn systems and T Tauri stars. 
For each processed source, the {\tt rot\_mod} pipeline provides a set of 
parameters characterising stellar rotation and magnetic activity, together with 
a list of transit identifiers corresponding to photometric outliers in the 
\gaia\ time series. A transit corresponds to the passage of a source across the 
entire \gaia\ focal plane, during which the satellite acquires quasi-simultaneous 
measurements in the broad-band $G$, \bp, and \rp\ passbands.

In this work, the transits flagged as photometric outliers are interpreted as
candidate flare events. Overall, the {\tt rot\_mod} pipeline identified
3\,314\,869 such outlier transits in 472\,963 stars, which constitute the
initial set of candidates analysed in this study.

\subsection{Gaia photometric time series}

The photometric time series used in this analysis are the $G$, \bp, and \rp\ 
epoch photometry available in \gaia\ DR3. The time series were cleaned following the 
procedure described in \citet{2023A&A...674A..13E}.

\FloatBarrier

\section{\label{sec:method}The method}

The sparse temporal sampling of the \gaia\ mission does not allow us to track
the temporal evolution of a flare event in detail. In the \gaia\
photometric time series, flares appear as individual transits in which the flux
is enhanced and the source becomes bluer with respect to its average flux and
colour.

\begin{figure*}[!t]
\begin{center}
\includegraphics[width=150mm]{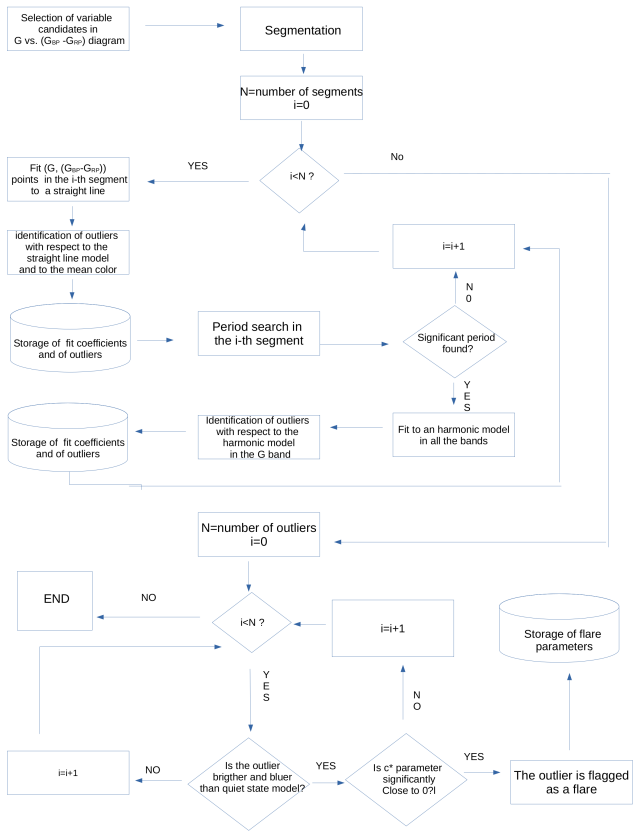}
\caption{ Flow diagram of the \gaia processing chain relevant to this work. The first block summarises the main steps of the {\tt rot\_mod} pipeline, which identifies photometric outliers and provides the initial list of candidate flare events. The subsequent blocks correspond to the { \tt flaring\_stars} pipeline presented in this work, which rejects spurious detections and flags the remaining candidates as confirmed flares.} 
\label{fig:diagram}
\end{center}
\end{figure*}

In this section we describe the pipeline adopted within the \gaia\ DPAC
(Data Processing and Analysis Consortium) to detect and characterise flare
events in the \gaia\ photometric time series. The processing chain relevant to
this work consists of two sequential components: the {\tt rot\_mod} module and
the {\tt flaring\_stars} module (Fig.~\ref{fig:diagram}).

The two modules perform distinct tasks. As discussed in Sect.~2.1, the {\tt rot\_mod} module measures the
stellar rotation period and identifies photometric outliers in the time series,
which constitute a list of {\it candidate flare events}. These candidates are
not yet validated and may include both genuine flares and spurious measurements.
The {\tt flaring\_stars} module subsequently analyses these candidates to
confirm real flare events and reject spurious detections.

The {\tt rot\_mod} module has been described in detail by
\citet{2023A&A...674A..20D}. Its main features are briefly summarised below,
while the {\tt flaring\_stars} module is presented here for the first time.

\subsection{Candidate selection}

The candidate selection is performed by the \gaia\ {\tt rot\_mod} pipeline,
whose main steps are summarised in the first block of Fig.~\ref{fig:diagram}.
Briefly, the pipeline selects stars that can exhibit magnetic activity based on
their position in the $M_G\text{-}(G_{\rm BP}-G_{\rm RP})$ colour--magnitude
diagram.

Once a source has been selected, the pipeline segments the \gaia\ time series
into overlapping intervals not exceeding 120~d and searches for rotational
modulation signals and flare candidates in each segment. As discussed in
\cite{2023A&A...674A..20D}, this segmentation is required because the long-term
evolution of spots and faculae affects the amplitude and shape of the rotational
modulation signal and could hamper its detection.

The {\tt rot\_mod} pipeline selects candidate flare events in two successive
steps.

In the first step, the pipeline performs, within each segment, a robust linear
regression between the $G$ magnitude and the $(G_{\rm BP}-G_{\rm RP})$ colour
and computes the mean colour of the segment. It then identifies all transits
significantly deviating either from the regression model or from the mean
colour, that is, those satisfying at least one of the following conditions:

\begin{equation}
\label{crit1}
|\epsilon_{{\rm slm},i}| >
\overline{|\epsilon_{\rm slm}|} + 3\,\sigma_{|\epsilon_{\rm slm}|}
\end{equation}

\begin{equation}
\label{crit2}
|(G_{\rm BP}-G_{\rm RP})_i - \overline{(G_{\rm BP}-G_{\rm RP})}|
> 3\,\sigma_{(G_{\rm BP}-G_{\rm RP})}
\end{equation}

These outliers are flagged as candidate flare events and temporarily removed
from the time series for the purpose of the period search, in order to allow a
reliable detection of the rotational modulation signal in the cleaned data.

In the second step, the pipeline performs a period search in each cleaned
segment. If a significant period is detected, the segment is further analysed
and its data points are fitted with the harmonic model:

\begin{equation}
\label{model}
m(t)=a_{m} + b_{m}\cos\left(\frac{2\pi t}{P}\right) + c_{m}\sin\left(\frac{2\pi t}{P}\right),
\end{equation}

where $P$ is the detected period and $m \in (G, G_{\rm BP}, G_{\rm RP})$.

After the fitting procedure, the pipeline computes the residuals of the $G$-band
data points from the harmonic model and identifies those satisfying:

\begin{equation}
\label{crit3}
|{\epsilon_{\rm hm}}_{i}| > \overline{|\epsilon_{\rm hm}|} + 3\sigma_{|\epsilon_{\rm hm}|}
\end{equation}

Data points satisfying Eq.~\ref{crit3} are also flagged by the {\tt rot\_mod}
pipeline as candidate flare events. Therefore, the final candidate list produced
by {\tt rot\_mod} includes all outliers identified through
Eqs.~\ref{crit1}--\ref{crit3}. These events are stored in the main database
together with their transit IDs and occurrence times.

At this stage, the selection is intentionally inclusive: the candidate set may
contain genuine flares, spurious events, and also outliers that are dimmer or
redder than the average magnitude and colour of the segment. These events are
therefore not yet considered confirmed flares.

In Fig.~\ref{example} we show the illustrative case of the star
{\tt Gaia DR3 2925085041699059712}. The full $G$-band time series (top-left
panel) exhibits a min-to-max amplitude of about 0.15~mag, mainly due to
long-term changes in the stellar magnetic field configuration affecting the
distribution of spots and faculae over the stellar surface. These variations
occur on timescales of years and resemble quasi-periodic magnetic cycles
observed in the Sun and in many magnetically active late-type stars
\citep[see e.g.][]{2017A&A...606A..58D,2016A&A...590A.133O}

A shorter interval of the same time series (top-right panel) reveals the
rotational modulation signal, characterised by a period $P \simeq 6$~d and an
amplitude of about 0.05~mag. The red symbols in both panels mark the transits
detected as candidate flare events by the pipeline. The horizontal black
segments in the top-left panel indicate the 120-day intervals used for the
analysis, while the green-shaded regions highlight the segments in which
candidate flares were identified.

The middle panels of Fig. \ref{example}  show the $G\text{-}(G_{\rm BP}-G_{\rm RP})$ diagrams for the
two segments, with the best-fitting straight lines over-plotted. The bottom
panels display the time--colour plane $(t, G_{\rm BP}-G_{\rm RP})$ for each
segment. The black lines represent the mean colour
$\overline{(G_{\rm BP}-G_{\rm RP})}$ and the red lines indicate the $3\sigma$
thresholds adopted to detect potential flare candidates.

In the middle and
bottom panels, the red points mark the candidate events detected through
Eqs.~\ref{crit1} and \ref{crit2}, whereas the cyan point marks the F3 event,
which is not detected by either of these two criteria. This event is identified
as an outlier only after applying Eq.~\ref{crit3} to the residuals of the
harmonic fit. In the top panel of Fig.~\ref{amp5}, we show the $G$-band measurements collected
in the fifth segment, folded according to the detected period and over-plotted
with the best-fitting harmonic model. In this case, the criterion given by
Eq.~\ref{crit3} permits the detection of the F3 event.
\subsection{Validation of the candidate flare events}

\begin{figure*}
\begin{center}
\includegraphics[width=160mm]{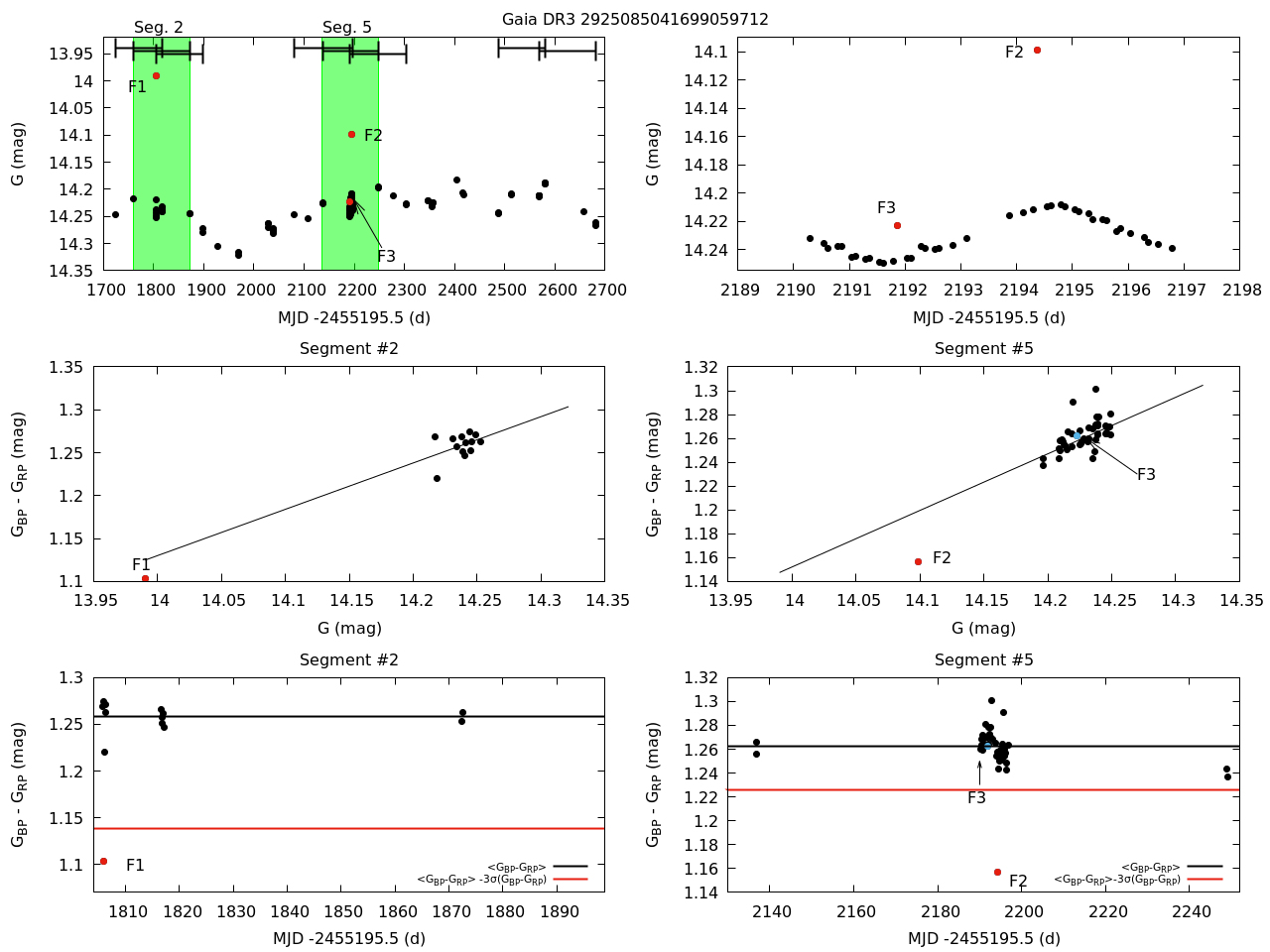}
\caption{Top-left panel: full {\it Gaia} $G$ time series for the star
{\tt Gaia DR3 2925085041699059712}. The dark horizontal lines indicate all
120-day segments into which the time series was divided by the pipeline for flare
searching. The green-shaded regions highlight the two segments (numbers~2 and~5)
where flare candidates were detected. Adjacent segments partly overlap
\citep[see][for details]{2023A&A...674A..20D}. Top-right panel: time-series slice
in which the variability resulting from rotational modulation is visible.
Middle-left panel: $G\text{-}(G_{\rm BP} -G_{\rm RP})$ diagram for the measurements
collected in the second segment. The straight line best fitting the data is
over-plotted. Middle-right panel: same as the middle-left panel for the fifth
segment. Bottom-left panel: $(t,\,G_{\rm BP}-G_{\rm RP})$ measurements collected
in the second segment. The black line marks the average colour in the segment.
The red line marks the threshold colour used to detect flare candidates.
Bottom-right panel: same as the bottom-left panel for the fifth segment. In all
panels the detected candidate flare events are labelled as $F1$, $F2$, and $F3$.}
\label{example}
\end{center}
\end{figure*}

\begin{figure}
\begin{center}
\includegraphics[width=80mm]{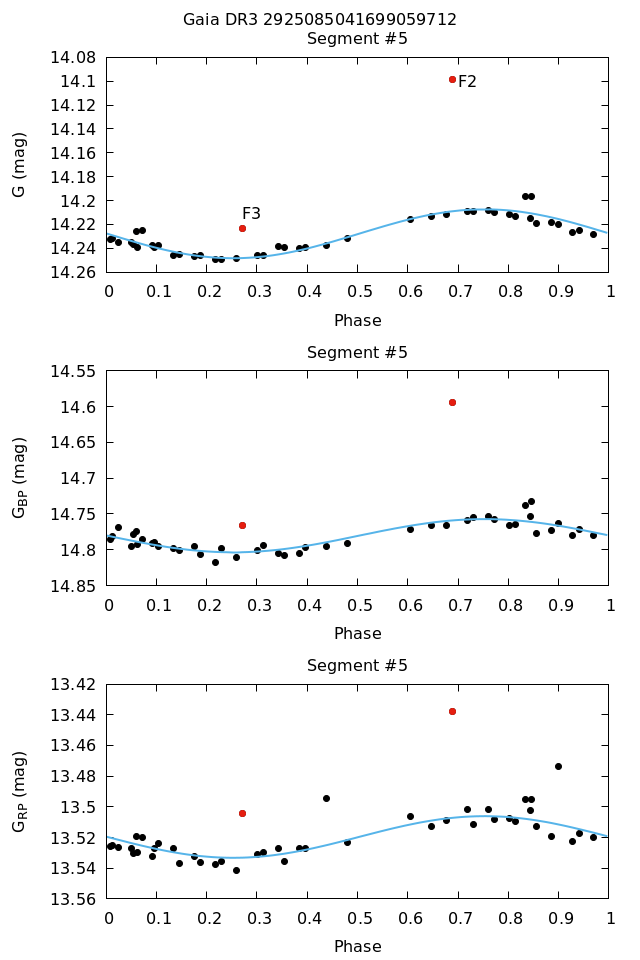}
\caption{$G$ (top panel), $G_{\rm BP}$ (middle panel), and $G_{\rm RP}$ (bottom
panel) measurements collected in the fifth segment. In all panels the points are
folded according to the stellar rotation period detected in the segment. The cyan
lines mark the harmonic functions best fitting the data and used to model the
quiet state of the star.}
\label{amp5}
\end{center}
\end{figure}

\begin{figure}
\begin{center}
\includegraphics[width=80mm]{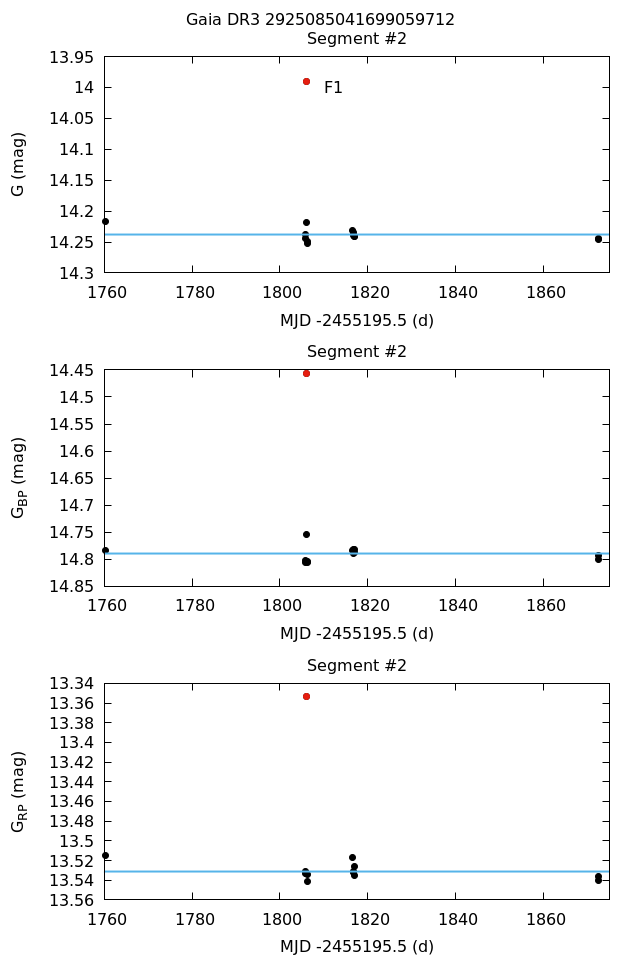}
\caption{$G$ (top panel), $G_{\rm BP}$ (middle panel), and $G_{\rm RP}$ (bottom
panel) measurements collected in the second segment. In all panels the cyan lines
mark the quiet-state magnitudes used to measure the flare amplitude.}
\label{amp2}
\end{center}
\end{figure}

The {\tt flaring\_stars} module processes the list of candidate flare events
identified by the {\tt rot\_mod} pipeline and performs a two-step validation
procedure to confirm genuine flares and reject spurious detections.

In the following, we denote with $G_{\rm out}$ and
$(G_{\rm BP}-G_{\rm RP})_{\rm out}$ the observed magnitude and colour of a
candidate event. Once an outlier satisfies the validation criteria and is
confirmed as a flare, we refer to the same observed quantities as
$G_{\rm flare}$ and $(G_{\rm BP}-G_{\rm RP})_{\rm flare}$.

In the first step, each candidate is tested against the expected behaviour of
a flare, namely a statistically significant brightening and simultaneous
blueing with respect to the stellar quiet state.

In order to establish whether a given outlier corresponds to a genuine flare
event, it is necessary to assess the stellar quiet state at the time of the
candidate flare. The quiet state is defined as the pair of $G$ and
$(G_{\rm BP}-G_{\rm RP})$ values that the star would exhibit in the absence of
flare activity.

Determining $G_{\rm quiet}$ and $(G_{\rm BP}-G_{\rm RP})_{\rm quiet}$ is not
straightforward because, as shown in Fig.~\ref{example}, the quiet state of a
magnetically active star varies in time owing to rotational modulation and to
the long-term evolution of spots and faculae.

To estimate $G_{\rm quiet}$ and $(G_{\rm BP}-G_{\rm RP})_{\rm quiet}$ we adopt
two different approaches.

If the outlier is detected in a segment for which the {\tt rot\_mod} pipeline
was able to fit the data with a harmonic model, the quiet state is defined as

\begin{equation}
\label{quietmodel_harmonic}
G_{\rm quiet}^{\rm hm} = G(t_{\rm out})
\end{equation}

\begin{equation}
(G_{\rm BP}-G_{\rm RP})_{\rm quiet}^{\rm hm} = G_{\rm BP}(t_{\rm out}) - G_{\rm RP}(t_{\rm out}),
\end{equation}
where $G(t_{\rm out})$, $G_{\rm BP}(t_{\rm out})$, and $G_{\rm RP}(t_{\rm out})$ are the values predicted by the harmonic model at the epoch of the outlier, $t_{\rm out}$.

If, on the other hand, the {\tt rot\_mod} module does not detect a significant
rotational modulation signal, the quiet values are approximated by first clipping the time-series segment, i.e. excluding all outliers, and then computing the mean:

\begin{equation}
\label{quietmodel_segment}
G_{\rm quiet}^{\rm mean} = \langle G_{{\rm seg},i} \rangle
\end{equation}

\begin{equation}
\label{quietcolorseg}
(G_{\rm BP}-G_{\rm RP})_{\rm quiet}^{\rm mean}=
\langle (G_{\rm BP}-G_{\rm RP})_{{\rm seg},i} \rangle
\end{equation}

The two approaches adopted to model the stellar quiet state are illustrated in
Figs.~\ref{amp5} and \ref{amp2} for the star
{\tt Gaia DR3 2925085041699059712}. Figure~\ref{amp5} shows the case in which
a harmonic model is used to describe the quiet state, whereas Fig.~\ref{amp2}
illustrates the case in which the quiet state is approximated by mean values in
the absence of a significant rotational modulation signal.

A candidate event is validated as a flare if it satisfies:

\begin{equation}
\label{critflare1}
G_{\rm out} - G_{\rm quiet} < -3 \sqrt{\sigma^2_{G_{\rm quiet}} + \sigma^2_{G_{\rm out}}}
\end{equation}

\begin{equation}
\label{critflare2}
(G_{\rm BP}-G_{\rm RP})_{\rm out} - (G_{\rm BP}-G_{\rm RP})_{\rm quiet}
< -3 \sqrt{
\sigma^2_{(G_{\rm BP}-G_{\rm RP})_{\rm quiet}} +
\sigma^2_{(G_{\rm BP}-G_{\rm RP})_{\rm out}}
}
\end{equation}

Applying these criteria reduces the initial set of candidates to 5\,172
validated flare events, which are consistent with the expected photometric
signature of stellar flares.

\subsection{Identification and removal of spurious events}

In the second validation step, the 5\, 172 events that passed the brightness and colour
criteria are further filtered to remove spurious detections caused by
instrumental or calibration issues.

In order to clean the sample, we employ the following criterion:

\begin{equation}
\label{clean}
|c^*(G_{\rm flare},(G_{\rm BP} - G_{\rm RP})_{\rm flare})|
< 3\sigma_{c^*}(G_{\rm flare})
\end{equation}

where $c^*_{\rm flare}$ is the per-transit corrected excess factor.

This parameter measures the consistency between the $G$, $G_{\rm BP}$, and
$G_{\rm RP}$ fluxes collected during a given transit. In the absence of
instrumental effects, these fluxes are expected to be mutually consistent.

\begin{figure}
\begin{center}
\includegraphics[width=80mm]{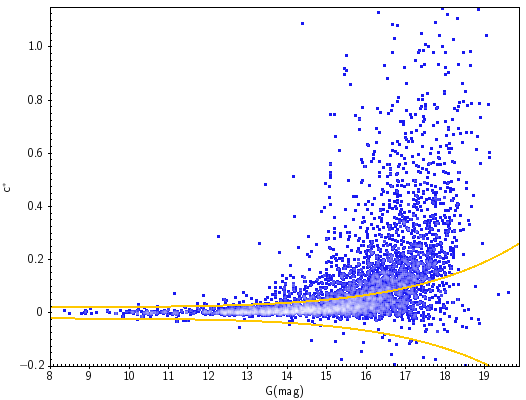}
\caption{Per-transit corrected excess flux of the candidate flare events vs. the stellar magnitude G}
\label{excessfactorg}
\end{center}
\end{figure}

In Fig.~\ref{excessfactorg} we show the $c^{*}$ parameter as a function of $G$,
and the yellow lines mark the adopted $3\sigma(G)$ thresholds.

Events lying outside these thresholds are rejected from the final sample.
After applying this criterion, the final sample consists of 3\,217 confirmed
flare events occurring in 2\,818 stars.

This step significantly reduces the number of validated events, highlighting
the importance of the excess-factor filtering in removing spurious detections.

\subsection{Measurement of flare amplitudes}

For each confirmed flare event, the {\tt flaring\_stars} pipeline measures the
flare amplitudes in the \gaia\ $G$, $G_{\rm BP}$, and $G_{\rm RP}$ bands, as
well as in the $(G_{\rm BP}-G_{\rm RP})$ colour. In all cases, the amplitudes
are computed with respect to the adopted quiet-state model, that is, either the
harmonic model when a significant rotational modulation signal is available, or
the clipped mean values otherwise. As illustrated in Figs.~\ref{amp2} and
\ref{amp5}, the adopted quiet-state model depends on the properties of the
segment in which the flare is detected.

The amplitudes are defined as:

\begin{equation}
A(G)= G_{\rm flare} - G_{\rm quiet}
\end{equation}

\begin{equation}
A(G_{\rm BP})= {G_{\rm BP}}_{\rm flare} - {G_{\rm BP}}_{\rm quiet}
\end{equation}

\begin{equation}
A(G_{\rm RP})= {G_{\rm RP}}_{\rm flare} - {G_{\rm RP}}_{\rm quiet}
\end{equation}

\begin{equation}
A(G_{\rm BP}- G_{\rm RP})_{\rm flare}=A(G_{\rm BP})_{\rm flare} - A(G_{\rm RP})_{\rm flare}
\end{equation}

When the quiet state is approximated by clipped mean values rather than by a
harmonic model, the resulting flare amplitudes are affected by larger
uncertainties, because the intrinsic stellar variability is not explicitly
modelled and removed from the time series.For this reason, the flare catalogue includes a flag indicating the method used to estimate the stellar quiet state. In particular, a value of 0 corresponds to amplitudes computed with respect to a mean model, while a value of 1 indicates that a harmonic model was adopted.

Finally, all amplitudes measured by the pipeline should be regarded as lower
limits to the true flare amplitudes. Owing to the sparse temporal sampling of
\gaia, the detected outlier may correspond to any phase of the flare evolution
(rise, peak, or decay), and the true maximum brightness increase may therefore
not be observed.

\subsection{Computation of the flares black-body temperatures}
\label{sec:tcomputation}

The multi-band photometry of \gaia\ enables a rough estimate of the flare black-body temperature from the ratio of the flux increases measured in different passbands.

Consider a star with effective temperature $T_{\rm eff}$, assumed to emit as a
black body $B(T_{\rm eff}, \lambda)$. The flux received from the star in a
given passband $X$ can be approximated as

\begin{equation}
\label{flux}
F_X = A_* \int_{0}^{\infty} B(T_{\rm eff}, \lambda)\, \psi_X(\lambda)\, d\lambda
\end{equation}

where $A_*$ is the projected area of the stellar disc and $\psi_X(\lambda)$ is
the transmittance of the passband. Limb darkening is neglected.

Now assume that a flare with projected area $A_F$ and effective temperature
$T_{\rm flare}$, also emitting as a black body, occurs on the stellar surface.
The flux received from the flaring star in passband $X$ is

\begin{equation}
\label{fluxflare}
F^*_X = \int_0^{\infty} \left[(A_* - A_F) B(T_{\rm eff}, \lambda) + A_F B(T_{\rm flare}, \lambda)\right] \psi_X(\lambda)\, d\lambda .
\end{equation}

The corresponding flux increase induced by the flare is therefore

\begin{equation}
\label{fluxincrease}
\Delta F_X = A_F \int_0^{\infty} \left[B(T_{\rm flare}, \lambda) - B(T_{\rm eff}, \lambda)\right] \psi_X(\lambda)\, d\lambda .
\end{equation}

We define the ratio of the flux increases in the $G_{\rm BP}$ and $G_{\rm RP}$ passbands as

\begin{equation}
\label{ratio}
Q = \frac{\Delta F_{G_{\rm BP}}}{\Delta F_{G_{\rm RP}}}
=
\frac{
\int_0^{\infty} \left[B(T_{\rm flare}, \lambda) - B(T_{\rm eff}, \lambda)\right] \psi_{G_{\rm BP}}(\lambda)\, d\lambda
}{
\int_0^{\infty} \left[B(T_{\rm flare}, \lambda) - B(T_{\rm eff}, \lambda)\right] \psi_{G_{\rm RP}}(\lambda)\, d\lambda
}.
\end{equation}

This ratio depends only on $T_{\rm flare}$ and $T_{\rm eff}$ and is independent
of the unknown flare area $A_F$, which cancels out.

\begin{figure}
\begin{center}
\includegraphics[width=80mm]{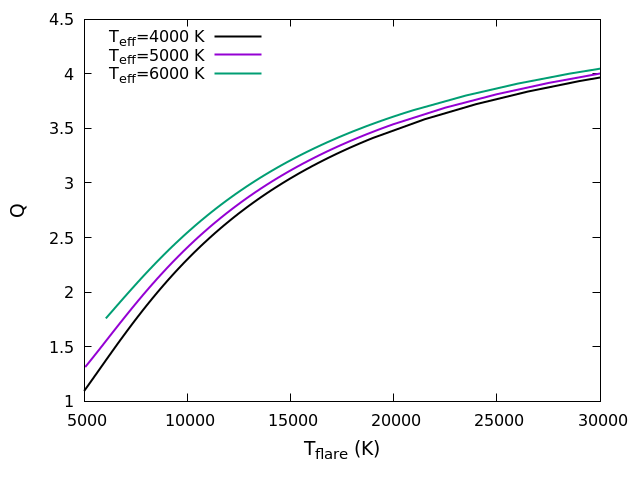}
\caption{Flare temperature $T_{\rm flare}$ as a function of the ratio $Q$ of the
flux increases measured in the $G_{\rm BP}$ and $G_{\rm RP}$ passbands. The
curves correspond to three representative stellar effective temperatures:
$T_{\rm eff} = 4\,000$, $5\,000$, and $6\,000$~K.}
\label{curva}
\end{center}
\end{figure}

Equation~\ref{ratio} allows us to tabulate the function $Q(T_{\rm flare})$ for a
given stellar effective temperature $T_{\rm eff}$. In
Fig.~\ref{curva} we show, as an example, the behaviour of $Q$ as a function of
$T_{\rm flare}$ for three representative values of $T_{\rm eff}$. These relations,
and analogous ones computed for other values of $T_{\rm eff}$, are used to infer
the flare temperature from the observed value of $Q$.

The stellar effective temperature adopted for each source was taken from the
\gaia\ DR3 table {\tt gaiadr3.astrophysical\_parameters} and derived by the
GSP-Phot module \citep{2023A&A...674A..26C}.

For each detected flare, the observed value of $Q$ was computed as

\begin{equation}
Q =
\frac{
F^{\rm obs}_{G_{\rm BP}}(t_{\rm flare}) - F^{\rm quiet}_{G_{\rm BP}}(t_{\rm flare})
}{
F^{\rm obs}_{G_{\rm RP}}(t_{\rm flare}) - F^{\rm quiet}_{G_{\rm RP}}(t_{\rm flare})
}.
\end{equation}

The fluxes used in Eq.~(20) are obtained by converting the observed \gaia\
magnitudes into physical fluxes using the photometric calibration of the
relevant \gaia\ data release. In the present work, the DR3 calibration,
relations, and passband definitions described in
\citet{2021A&A...649A...3R} are adopted, ensuring consistency with the response
functions used in Eq.~(\ref{flux}).

This procedure is affected by several sources of uncertainty:

\begin{itemize}
\item the uncertainty associated with the flare phase corresponding to the observed data point;
\item the uncertainty in the stellar effective temperature $T_{\rm eff}$;
\item the uncertainty in estimating the quiet-state magnitude.
\end{itemize}

The flare temperature evolves during the event, typically reaching its maximum
at peak brightness and decreasing during the decay phase
\citep[e.g.][]{2020ApJ...902..115H}. Because \gaia\ provides only single-epoch
measurements, the inferred temperature reflects the specific phase of the flare
at the time of observation and cannot be corrected for this effect.

The uncertainty in $T_{\rm eff}$ propagates directly into the estimated flare
temperature, since the flux ratio $Q$ depends on both $T_{\rm eff}$ and
$T_{\rm flare}$. The uncertainty in the quiet-state magnitude arises from the
adopted model for the quiescent flux level and from the photometric precision
of the data.

To limit the impact of the uncertainties related to $T_{\rm eff}$ and to the
estimation of the quiet-state magnitude, we computed $T_{\rm flare}$ only for
flares whose amplitudes were determined with respect to a harmonic model, since
only in this case the quiescent flux level can be reliably modelled (see
Sect.~3.2). This requirement restricts the sample to 486 events.

To further exclude temperature estimates affected by poor photometry, we
derived three independent estimates of $T_{\rm flare}$ using different
combinations of the \gaia\ passbands: ($G_{\rm BP}$, $G_{\rm RP}$),
($G$, $G_{\rm RP}$), and ($G$, $G_{\rm BP}$). We retained only those flares for
which the three estimates differ by less than 20\%, and for which the resulting
temperature is higher than the stellar effective temperature $T_{\rm eff}$.
  These combined criteria reduce the sample of flares with reliable $T_{\rm flare}$ estimates to 348 events. 
 
 For each of these events, the final $T_{\rm flare}$ value reported in Table~\ref{table:flares} was computed as the average of the three consistent estimates. This subset is not intended to be representative of the full flare sample.

\subsection{ Computation of the flare energy}
\label{sec:energy}
The \gaia\ sampling does not allow us to track the temporal evolution of a flare, estimate its duration, or integrate its total radiated energy. However, since the effective integration time of a single $G$-band transit is $t_{\rm int}=40$~s, it is possible to derive a lower limit to the flare energy.

However, considering that the integration time $t_{\rm int}$ for the flux in the G band is 40 seconds, it is possible to estimate a lower limit for the flare's energy $E_{\rm f,low}$ by employing the equation:

\begin{equation}
\label{lowerlimit}
E_{\rm{low}} = 4\pi D^{2} \, t_{\rm{int}} \, 
\left[ F^{\rm{obs}}_{G}(t_{\rm{flare}}) - F^{\rm{quiet}}_G(t_{\rm{flare}}) \right] \quad [\rm{erg}]
\end{equation}

where $F^{\rm{obs}}_{G}(t_{\rm{flare}}) - F^{\rm{quiet}}_G(t_{\rm{flare}}) $ is the flux increase induced by the flare in the $G$ band and $D$ is the distance of the star reported in the \gaia DR3 table {\tt gaiadr3.astrophysical\_parameters} and inferred from the GSP-Phot module. 
Distances are available only for a subset of the sample, as the GSP-Phot module provides solutions only for sources with reliable astrophysical parameter estimates.

The quantity$E_{\rm f,low}$   represents a lower bound for the energy released during the flare event, as it only accounts for the energy emitted within the integration time $t_{\rm int}$   and the wavelength range covered by the $Gaia$ $G$ passband, which spans from 300 to 1\,100 nm \citep[see][for further details on \gaia passbands]{2021A&A...649A...3R}

\subsection{Computation of $\rm H_\alpha$ excess}

The ${\rm H_\alpha}$ emission line is a well-known tracer of stellar magnetic
activity \citep[see, e.g.,][]{1979ApJ...234..579C,1995A&A...294..165M}.

In the \gaia\ context, an ${\rm H_\alpha}$-sensitive photometric index can be
derived from the mean $G_{\rm BP}$ and $G_{\rm RP}$ spectra using the GaiaXPy
tool\footnote{Available at \url{https://gaia-dpci.github.io/GaiaXPy-website/}}.
This tool defines the narrow-band photometric system
{\tt ELS\_custom\_w09\_s2}, in which the ${\rm H_\alpha}$ emission is quantified
through the colour index $m_{\rm H\alpha} - m_{\rm CH\alpha}$, where
$m_{\rm H\alpha}$ and $m_{\rm CH\alpha}$ are the magnitudes corresponding to two
photometric bands centred on the ${\rm H_\alpha}$ line and on the adjacent
continuum, respectively \citep[see][]{2023A&A...674A..33G}.

  The colour index $m_{\rm H\alpha} - m_{\rm CH\alpha}$ depends not only on magnetic activity but also on the stellar effective temperature and metallicity. To obtain a more reliable tracer of chromospheric activity, we introduce the ${\rm H_\alpha}$ colour excess, defined as
\begin{equation}
\label{halphaindex}
E(m_{\rm H\alpha}-m_{\rm CH\alpha}) = (m_{\rm H\alpha}-m_{\rm CH\alpha})_{\rm obs} - (m_{\rm H\alpha}-m_{\rm CH\alpha})_{\rm exp},
\end{equation}
where $(m_{\rm H\alpha} - m_{\rm CH\alpha})_{\rm obs}$ is the observed colour index and $(m_{\rm H\alpha} - m_{\rm CH\alpha})_{\rm exp}$ is the colour index expected for a magnetically inactive star with the same effective temperature and metallicity as the target. The procedure used to determine $(m_{\rm H\alpha} - m_{\rm CH\alpha})_{\rm exp}$  is explained in detail in Appendix A. 
By construction, this excess removes the dependence of the raw index on stellar
parameters, allowing a more direct comparison of chromospheric activity levels
across different stars.

A more negative colour excess corresponds to stronger ${\rm H_\alpha}$ emission
relative to the adopted inactive reference and, consequently, to higher levels
of stellar magnetic activity.
Conversely, positive values indicate that the observed
$m_{\rm H\alpha}-m_{\rm CH\alpha}$ colour lies above the adopted inactive
reference relation. These values are found for only a small fraction of the
sample and are typically very close to zero; we therefore do not assign them
a direct physical interpretation in terms of lower magnetic activity.

In some cases (e.g. young stellar objects), strong ${\rm H_\alpha}$ emission may
also be associated with accretion processes; however, this aspect is not further
investigated in the present work.

\section{Results}

We applied the {\tt rot\_mod}--{\tt flaring\_stars} pipeline to all the 474\,026 stars listed in the {\tt gdr3\_rotmod} catalogue. The pipeline identified 25\,080 flare candidates, of which 3\,217 passed the cleaning procedure defined by Eq.~\ref{clean}. 

Our final catalogue therefore contains 3\,217 flares detected in 2\,818 stars. Fourteen of these sources were already reported in previous works, whereas the remaining objects are newly identified flaring stars. Three of the flares detected here are also clearly visible in the {\it Kepler}/K2 light curves analysed by \citet{2021A&A...645A..42I} (see discussion in Appendix~\ref{sec:previousworks}). The flare temperature could be estimated for 348 of the detected events.

The global incidence of flaring stars in the analysed sample is therefore $\sim$0.6\%. 
This relatively low detection rate is mainly due to the sparse temporal
sampling of the mission. However, it increases by a factor of $\sim$3 towards
cooler stars, reaching values of $\sim$1.5--2\% in the M-dwarf regime (see
Sect.\ref{detrate}).

The results of this work are summarised in Tables~\ref{table:stars} and~\ref{table:flares}, which report the catalogues of the flaring stars and of the detected flares, respectively. For each table, only the first ten rows are shown here; the full versions are available in electronic form at the CDS (Centre de Donn\'ees astronomiques de Strasbourg).

In the following subsections we describe the physical properties of the flaring stars detected here, assess the sensitivity and completeness of the pipeline, and discuss the statistical properties of the flares and their relationship with the main stellar parameters.

\subsection{Properties of the flaring stars} 

\begin{figure*}[!t]
\begin{center}
\includegraphics[width=160mm]{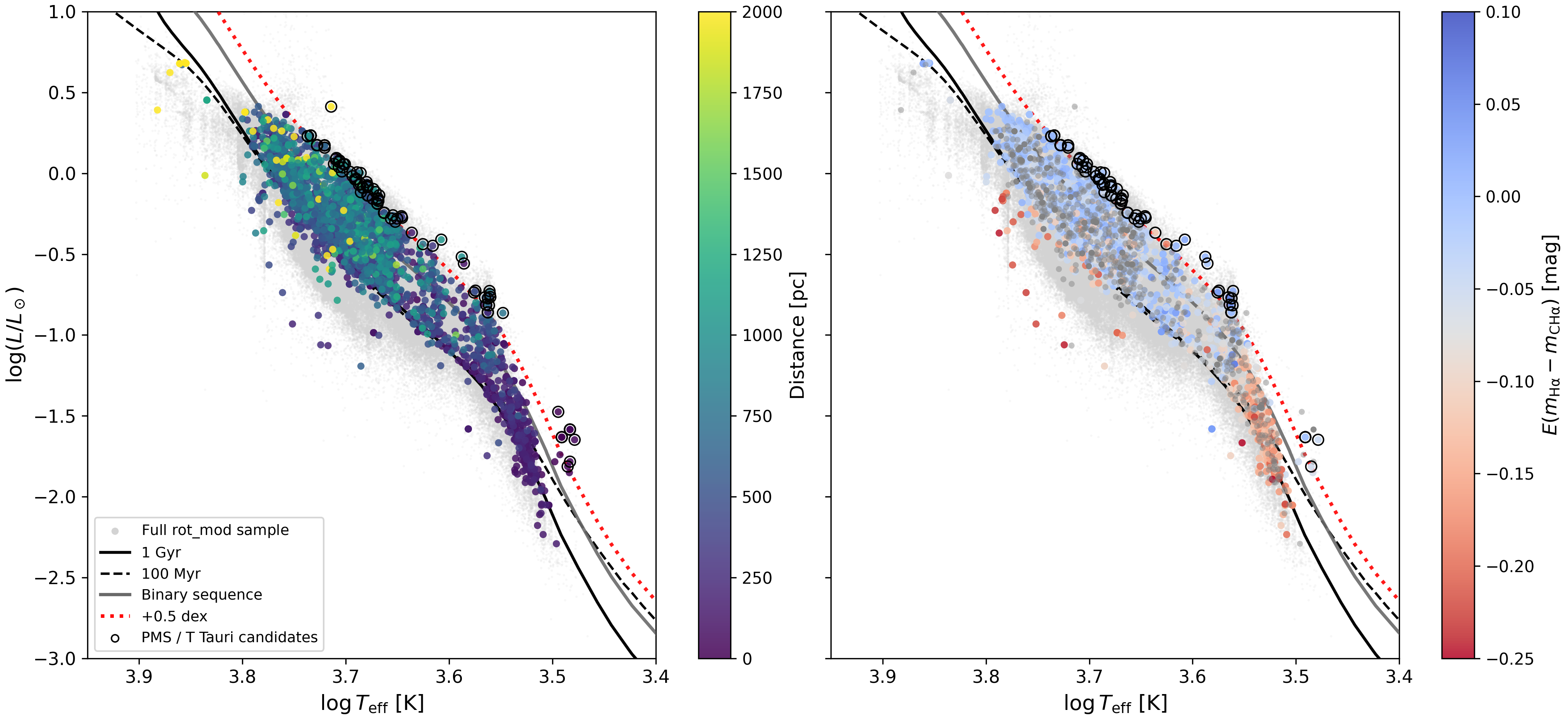}
\caption{Hertzsprung--Russell diagram of the flaring stars.
The full {\tt gdr3\_rotmod} parent sample is shown in light grey for comparison.
Left: flaring stars colour-coded by distance (in pc).
Right: flaring stars colour-coded by $\rm H_{\alpha}$ colour excess.
Negative values indicate stronger H$_\alpha$ emission
relative to the inactive reference.
The solid and dashed black lines show the PARSEC isochrones for 1 Gyr and
100 Myr, respectively. The grey line marks the binary sequence, and the red
dotted curve indicates the 0.5 dex threshold used to identify T-Tauri
candidates (open circles).}
\label{hrdouble}
\end{center}
\end{figure*}

Figure~\ref{hrdouble} shows the distribution of the flaring stars in the
Hertzsprung–Russell (HR) diagram, colour-coded by distance (left panel) and by
the ${\rm H_\alpha}$ colour excess (right panel). For comparison,
the full {\tt gdr3\_rotmod} parent sample is also shown in light grey.
The effective temperature, luminosity, and distance values used to build the
diagram are taken from the \gaia\ DR3 table
{\tt gaiadr3.astrophysical\_parameters} and inferred by the GSP-Phot (General
Stellar Parametrizer from Photometry) module \citep[see][for
details]{2023A&A...674A..26C}.

The black solid and dashed curves represent the PARSEC 1 Gyr and 100 Myr
isochrones, respectively, while the grey solid line marks the binary sequence
and the red dotted line indicates the locus 0.5 dex above the main sequence,
adopted to identify pre-main-sequence (PMS) candidates.

Most flaring stars cluster close to the main sequence, as expected for
magnetically active late-type dwarfs. A fraction of sources is found above the
main sequence and extends up to and beyond the binary sequence, potentially
indicating unresolved binaries or stars with overestimated luminosities due to
uncertainties in distance or extinction. A smaller fraction
(\textasciitilde 3--4\%) lies above the 0.5 dex threshold, suggesting a
pre-main-sequence nature. The left panel shows that these PMS candidates are
typically located within $\sim$1 kpc, whereas the coolest flaring stars
($\log T_{\rm eff} \lesssim 3.55$) are found within 500 pc, consistent with
nearby young stellar populations, although this trend is partly influenced by
the detection limits for intrinsically faint objects.

A comparison with the parent sample reveals that, while the {\tt rot\_mod} stars
are distributed on both sides of the main sequence, the flaring stars are
preferentially located above it. Assuming that flaring stars trace the most
magnetically active component of the sample, this behaviour is consistent with
the findings of \citet{2023AJ....166...63J}, who showed that more active stars
tend to be displaced above the main sequence in Gaia colour--magnitude diagrams.

In the right panel, stars with large negative
$E(m_{\rm H\alpha}-m_{\rm CH\alpha})$ values, corresponding to stronger
${\rm H_\alpha}$ emission relative to the inactive reference, are
preferentially found among the coolest sources.

A small number of objects are found significantly below the main sequence.
The most strongly underluminous sources also show large
negative $E(m_{\rm H\alpha}-m_{\rm CH\alpha})$ values. A possible explanation
is that both their anomalous position in the HR diagram and their large
negative ${\rm H_\alpha}$ colour excesses are affected by inaccuracies in the
inferred stellar parameters, particularly $T_{\rm eff}$. An overestimated
$T_{\rm eff}$ would place these sources artificially towards the hot side of
the HR diagram and could therefore contribute to their apparent displacement
below the main sequence. At the same time, the expected
$m_{\rm H\alpha}-m_{\rm CH\alpha}$ colour used to compute the excess is
derived from an empirical relation that depends on $T_{\rm eff}$ and
metallicity. Uncertainties in these parameters, as well as in the reference
relation itself, therefore propagate into the derived ${\rm H_\alpha}$ colour
excess. The large negative
excesses of these sources should therefore be interpreted with caution.

\begin{figure}
\begin{center}
\includegraphics[width=80mm]{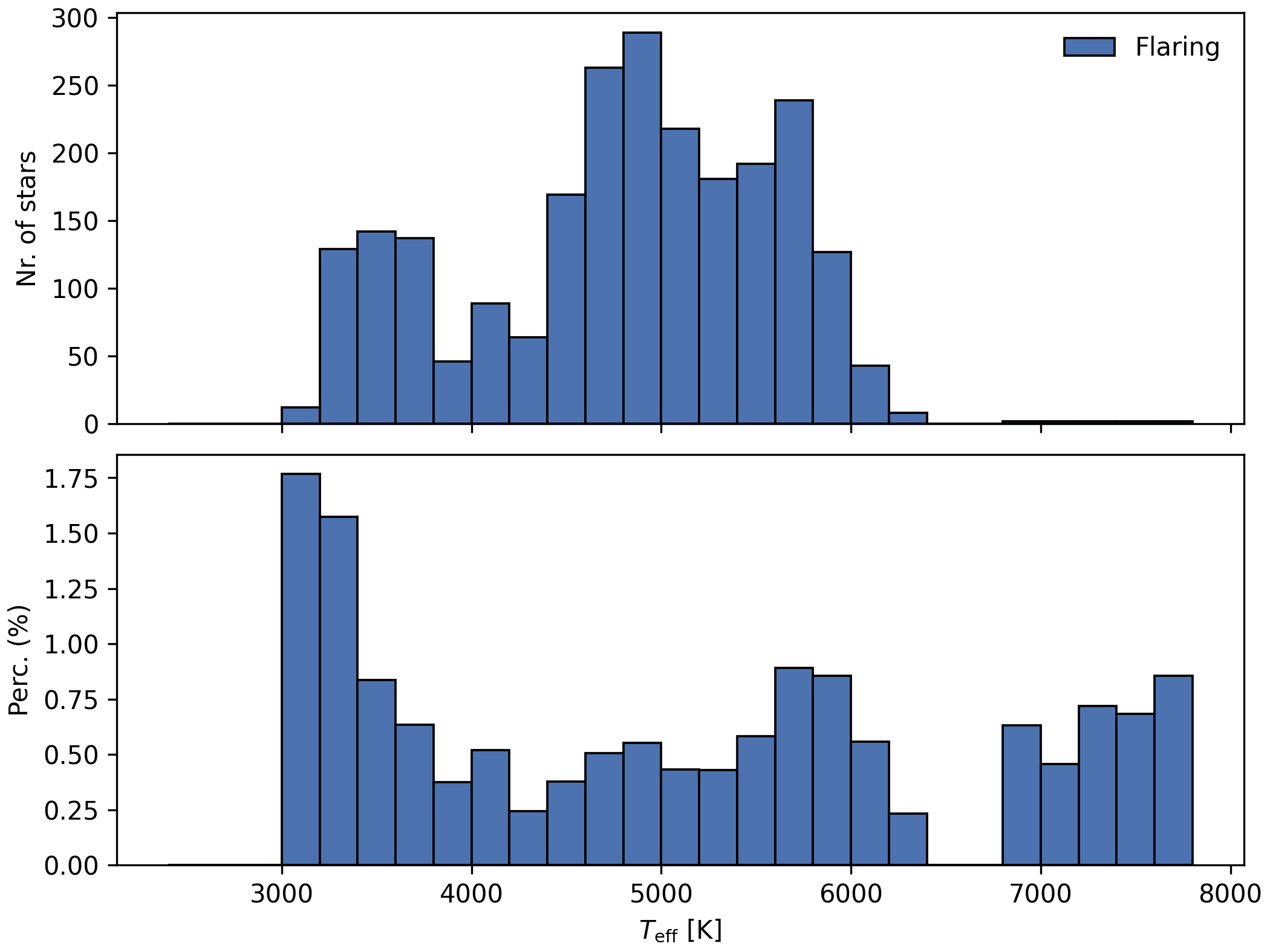}
\caption{Top panel: $T_{\rm eff}$ distribution of stars with detected flares.
Bottom panel: detection rate of flaring stars as a function of $T_{\rm eff}$,
defined as the percentage of stars with at least one detected flare in each
temperature bin.}

\label{disttargets}
\end{center}
\end{figure}

\subsection{Detection rate}
\label{detrate}

In Fig.~\ref{disttargets} we show the $T_{\rm eff}$ distribution of the flaring
stars (top panel) and the detection rate as a function of $T_{\rm eff}$ (bottom
panel). The detection rate is defined as the fraction of stars with at least one
detected flare relative to the total number of analysed stars in each temperature
bin, and   is expressed in percent in the figure.

The $T_{\rm eff}$ distribution of flaring stars peaks around
$T_{\rm eff}\sim 5\,000$~K, reflecting the underlying distribution of the
analysed stellar sample. In contrast, the detection rate increases towards
cooler temperatures and reaches a maximum of $\sim$1.8\% at
$T_{\rm eff}\sim 3\,000$~K.

This difference arises because the detection rate measures the relative
incidence of flares, which is intrinsically higher in late-type dwarfs, and
because flares are more easily detectable in cool stars, where the contrast
between the flare emission and the stellar photosphere is larger.

This behaviour is consistent with previous findings. \citet{2020AJ....159...60G}, based on the First TESS Data Release, showed that M-type stars flare more frequently than K-type stars and that, among M stars, the detection rate increases from spectral type M0 to M6. A similar trend was reported by \citet{2011AJ....141...50W}. 

The increasing detection rate towards late-type dwarfs confirms that the \gaia\ pipeline is sensitive to the enhanced flare activity of cool M stars, despite their intrinsic faintness.

\subsection{Detection completeness and sensitivity limits} 

\begin{figure}
\begin{center}
\includegraphics[width=80mm]{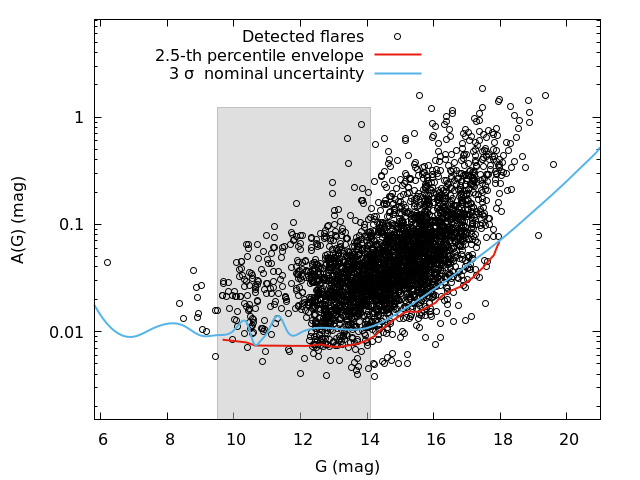}
\caption{Flare amplitudes as a function of stellar $G$ magnitude. 
The red solid line shows the empirical detection limit derived from the 2.5th percentile of the amplitude distribution, while the cyan dashed line represents three times the nominal \gaia single-transit uncertainty. 
The shaded region marks the magnitude interval of nearly uniform detection sensitivity.}
\label{sensitivity}
\end{center}
\end{figure}

The sensitivity of the {\tt flaring\_stars} pipeline was assessed by analysing the relation between flare amplitude and apparent magnitude. Figure~\ref{sensitivity} shows the amplitudes of the detected flares as a function of the stellar $G$ magnitude, together with the empirical detection limit (red solid line) and the expected noise level derived from the nominal \gaia\ photometric performance (cyan dashed line).

The empirical detection limit was derived from the observed amplitude distribution as a function of magnitude. The sample was sorted by $G$ and divided into bins containing 150 flares each. For every bin, we computed the 2.5th percentile of the flare amplitude distribution, which provides an estimate of the minimum detectable amplitude at that brightness. The resulting sequence was smoothed in logarithmic amplitude using a Savitzky–Golay filter and interpolated over a regular $G$ grid. The smoothed 2.5th-percentile envelope defines the empirical sensitivity curve. The reference amplitude, $A_{\rm ref}$, corresponds to the minimum of this curve and identifies the point of maximum sensitivity.

For comparison, the cyan dashed curve in Fig.~\ref{sensitivity} shows three times the nominal \gaia\ single-transit uncertainty converted into magnitudes. Both curves trace the increase of the noise floor with magnitude; however, the empirical limit lies systematically below the nominal $3\sigma$ threshold. This indicates that the effective sensitivity of the pipeline is slightly better than expected from the nominal single-transit photometric precision. The improvement arises mainly from the adopted estimate of the quiescent flux, which relies on multiple measurements and, when available, on a harmonic model that accounts for intrinsic stellar variability.

At bright mag ($G \lesssim 12$), the empirical envelope shows a mild increase in the minimum detectable amplitude. This behaviour is consistent with the instrumental configuration of \textit{Gaia}, which switches between different gates and window classes to prevent CCD saturation \citep[see][]{2021A&A...649A...2L,2021A&A...649A...3R}. While these transitions are reflected in the nominal uncertainty curve, the empirical envelope smooths out small-scale features owing to the limited number of bright sources in the sample.

The shaded region in Fig.~\ref{sensitivity} marks the magnitude interval where the empirical envelope remains within 10\% of its minimum value, corresponding to nearly uniform detection sensitivity. The minimum amplitude, $A_{\rm ref} \simeq 0.007$~mag, occurs at $G \simeq 13$ and represents the highest sensitivity achieved by the pipeline. The interval $9.5 \lesssim G \lesssim 14.1$ is therefore adopted as the completeness range in the following analysis.

\begin{table*}
\caption{Flaring stars catalogue}             
\label{table:stars}      
\centering          
\begin{tabular}{l l l l l l l }     
\hline\hline       
                     
\gaia DR3 sourceid & $T_{\rm eff}$  & ${\rm log} g$ & [{\rm M}/{\rm H}] & $E(m_{\rm H\alpha} -m_{\rm CH\alpha})$ & $P_{\rm rot}$ &$N_{\rm flares}$\\
& K&[dex] &[dex]  &[mag]&[d]&\\
\hline  
11980420432272896 & 4\,373 & 4.31 & -0.39 & -0.07 & 7.25 & 1\\
  11987807776011520 & 4\,904 & 4.27 & -1.48 & 0.01 & 2.04 & 1\\
  39115645852216192 & 4\,675 & 4.50 & -0.72 &  & 18.15 & 2\\
  39307510630668544 & 5\,017 & 4.65 & -0.68 &  & 1.01 & 2\\
  46622664569875840 & 4\,672 & 4.26 & -0.56 &  & 0.57 & 2\\
  50905056201179520 & 5\,849 & 4.35 & -0.34 & -0.01 & 2.48 & 2\\
  51112760819821568 &  &  &  &  & 1.33 & 1\\
  51526417710225408 & 3\,617 & 4.60 & -0.28 & -0.10 & 0.41 & 1\\
  52887303867228032 & 4\,641 & 4.39 & -1.28 & -0.07 & 0.44 & 1\\
  63838886357191936 & 3\,760 & 4.69 & 0.05 &  & 1.68 & 1\\
\hline                  
\end{tabular}\tablefoot{
$T_{\rm eff}$, ${\rm log} g$, $[\rm{M/H}]$ are the {\tt teff\_gspphot}, {\tt logg\_gspphot}, {\tt mh\_gspphot} parameters reported in the {\tt gaiadr3.astrophysical\_parameters} table. $P_{\rm rot}$ is the {\tt bestprot} parameter  taken from the  {\tt gaiadr3.vari\_rotation\_modulation} table. $N_{\rm flare}$ is the number of  detected flares. 
}
\end{table*} 

\begin{table*}
\caption{Flare parameters}             
\label{table:flares}      
\centering          
\begin{tabular}{l l l l l l l l }     
\hline\hline       
                     
\gaia DR3 Sourceid & MJD -2455195.5 \tablefootmark{a} & $A(G)$ & $A(G_{\rm BP})$ & $A(G_{\rm RP})$ & $E_{\rm f,low}$ &$T_{\rm flare}$ & $model_{\rm flag}$ \\
&   [d]& [mag]&[mag]&[mag]&[erg]&[K]& \\
\hline   
11980420432272896 & 1\,710.8632 & 0.064 & 0.036 & 0.032 & $7.77\times 10^{32}$& & 0 \\
11987807776011520 & 1\,827.1123 & 0.029 & 0.057 & 0.024 & $5.19\times 10^{32}$& & 0 \\
39115645852216192 & 2\,599.007 & 0.059 & 0.075 & 0.05 & $5.99\times 10^{32}$& & 1 \\
39115645852216192 & 2\,576.63 & 0.054 & 0.059 & 0.043 & $5.48\times 10^{32}$& & 1 \\
39307510630668544 & 1\,913.8108 & 0.043 & 0.12 & 0.048 & $1.16\times 10^{32}$& & 0 \\
39307510630668544 & 1\,913.9869 & 0.044 & 0.097 & 0.033 & $1.18\times 10^{32}$& & 0 \\
46622664569875840 & 1\,922.4926 & 0.036 & 0.031 & 0.028 & $8.74\times 10^{32}$& & 0 \\
46622664569875840 & 1\,914.4885 & 0.029 & 0.03 & 0.022 & $7.01\times 10^{32}$& & 0 \\
50905056201179520 & 1\,876.345 & 0.026 & 0.014 & 0.014 & $2.01\times 10^{33}$ & & 0 \\
50905056201179520 & 1\,839.3002 & 0.01 & 0.011 & 0.008 & $7.40\times 10^{32}$& 7091 & 0 \\
\hline                  
\end{tabular}
\tablefoot{
\tablefoottext{a}{Time at which the flare was detected.}\\
$model_{\rm flag}$: 0 = mean model, 1 = harmonic model.
}
\end{table*}

\begin{table*}[ht]
\centering
\caption{Spearman rank correlation coefficients ($r_s$) and associated $p$-values for the relationships between flare parameters and stellar parameters.  }
\label{spearmancoeff}
\begin{tabular}{llcccccc}
\hline
Flare parameter & Stellar parameter & $r_s$ & $p$ & $N$ & $r_s$ & $p$ & $N$ \\
 &  & \multicolumn{3}{c}{FS} & \multicolumn{3}{c}{CLS} \\
\hline
$A(G)$ & $P_{\rm rot}$ & $-0.24$ & \quad $<10^{-6}$ & 3\,217 & -0.20 &\quad $<10^{-6}$ & 2\,116\\
$A(G)$ & $T_{\rm eff}$ & $-0.43$ & \quad $<10^{-6}$ &2\,681 & -0.19 & \quad $<10^{-6}$& 1\,739 \\
$A(G)$ & $E(m_{\rm H\alpha} - m_{\rm CH\alpha})$ & $-0.40$ &\quad $<10^{-6}$ & 2\,005 & -0.23& \quad $<10^{-6}$ & 1\,507 \\
$T_{\rm flare}$ & $P_{\rm rot}$ & $0.11$ &\quad $0.035$ &348 & & & \\
$T_{\rm flare}$ & $T_{\rm eff}$ & $0.12$ &\quad $0.019$ &348 & & & \\
$T_{\rm flare}$ & $E(m_{\rm H\alpha} - m_{\rm CH\alpha})$ & $-0.10$ &\quad $0.11$ &269 & & & \\
\hline
\end{tabular}
\tablefoot{Values are reported for the full sample (FS) and for the completeness-limited sample (CLS), where applicable. The number of events ($N$) used in each correlation varies depending on the availability of the relevant stellar and flare parameters. For $A(G)$, the CLS corresponds to the magnitude range defined in Sect.~4.3. For $T_{\rm flare}$, correlation coefficients are computed only for the full sample, which consists of the subset of events with reliable temperature estimates.}
\end{table*}

\begin{figure*}[!t]
\begin{center}
\includegraphics[width=160mm]{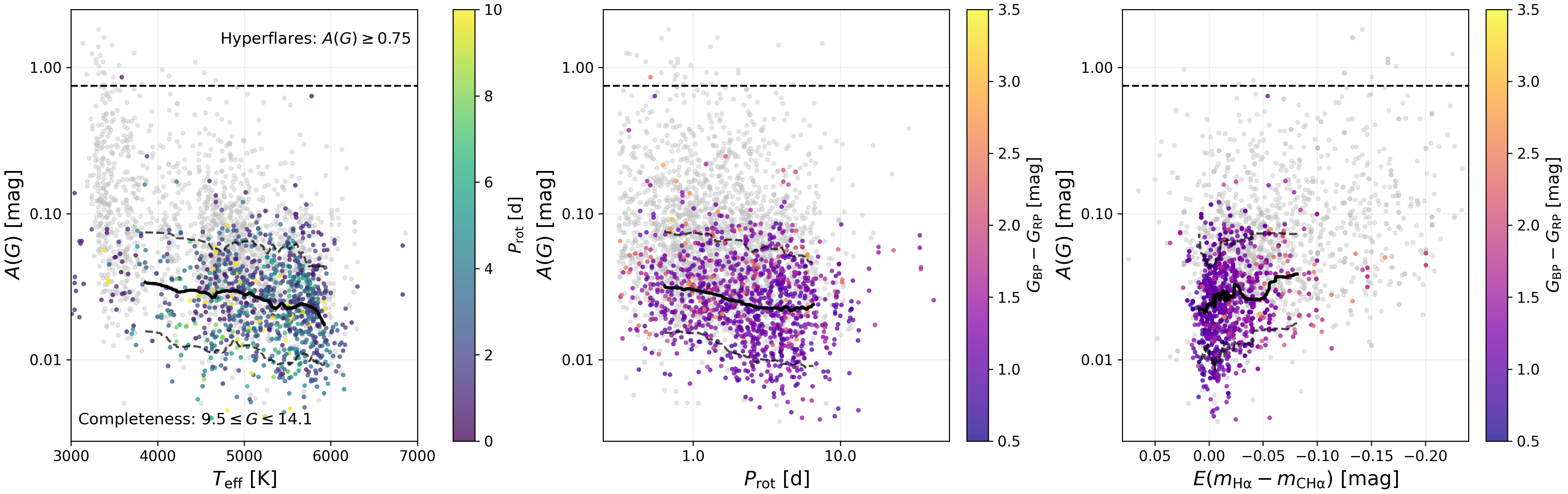}

\caption{Flare amplitude in the \gaia\ $G$ band, $A(G)$, as a function of
stellar effective temperature (left), rotation period (centre), and
${\rm H_\alpha}$ colour excess (right). Points within the completeness range
($9.5 \le G \le 14.1$) are colour-coded according to the stellar rotation period in the left panel and according to the $(G_{\rm BP}-G_{\rm RP})$ colour in the centre and right panels. Points outside this range are shown in light grey for reference. 
The solid black curves represent the running median computed from the completeness-limited sample, while the dashed black curves indicate the 10th and 90th percentiles of the amplitude distribution. The horizontal dashed line marks the hyper-flare threshold at $A(G)=0.75$ mag.}

\label{ampvs.par}
\end{center}
\end{figure*} 

\subsection{Correlation between flare amplitudes and stellar parameters}

Figure~\ref{ampvs.par} illustrates the behaviour of the flare amplitude in the
\gaia\ $G$ band, $A(G)$, as a function of stellar effective temperature
$T_{\rm eff}$ (left panel), rotation period $P_{\rm rot}$ (centre panel), and
${\rm H_\alpha}$ colour excess (right panel). In the left panel, flares within
the completeness range are colour-coded according to the stellar rotation
period, while in the centre and right panels they are colour-coded according to
the $(G_{\rm BP}-G_{\rm RP})$ colour. Points outside the completeness range
($9.5 \le G \le 14.1$; Sect.~4.3) are shown in light grey for reference.

Given the large intrinsic scatter of flare amplitudes and the
magnitude-dependent detection thresholds, we refrain from fitting analytical
relations to the data. Instead, to guide the eye, we overplot in each panel the
running median of $A(G)$ computed from the completeness-limited sample (solid
black lines). The dashed black curves indicate the 10th and 90th percentiles of
the amplitude distribution, providing a direct estimate of the dispersion.
A visual inspection of the running median suggests the presence of weak
systematic variations, with slightly larger flare amplitudes occurring in
cooler stars, shorter rotation periods, and stronger ${\rm H_\alpha}$ emission.
These variations, however, remain shallow and are dominated by a large intrinsic
dispersion.

To quantify the relations between $A(G)$ and the stellar parameters, we
computed the Spearman rank correlation coefficients for both the full sample
and the completeness-limited sample. The resulting coefficients and associated
$p$-values are reported in Table~\ref{spearmancoeff}.

A comparison between the two samples reveals a significant difference in the
inferred correlations. In the full sample, moderate correlations are found
between $A(G)$ and both $T_{\rm eff}$ and ${\rm H_\alpha}$ colour excess
(with $r_s \sim -0.4$). However, these correlations are strongly affected by
observational biases related to the magnitude-dependent detection threshold.

When restricting the analysis to the completeness-limited sample, the
correlations become significantly weaker ($|r_s| \lesssim 0.2$). This is
consistent with Fig.~\ref{ampvs.par}, where the running median highlights only
shallow variations combined with a large intrinsic dispersion of flare
amplitudes at fixed stellar parameters. Given the large number of detected
flares, even weak correlations yield very small $p$-values; however, the modest
values of $r_s$ indicate that these relationships are intrinsically weak.

The weak dependence of flare amplitude on $T_{\rm eff}$ suggests that stellar
effective temperature alone is not a strong predictor of flare amplitude. The
stronger relations observed in the full sample can be largely explained by
detection biases. In particular, flares on cool stars exhibit a higher contrast
against the stellar photosphere, which enhances their observed amplitudes
\citep[e.g.][]{2020AJ....159...60G}. On the other hand, flares on hotter stars
benefit from a higher signal-to-noise ratio due to their intrinsic brightness,
which favours the detection of low-amplitude events
\citep[e.g.][]{2011AJ....141...50W}. The interplay between these effects can
produce apparent correlations between flare amplitude and stellar effective
temperature. When these biases are mitigated by restricting the analysis to the
completeness-limited sample, the dependence of $A(G)$ on $T_{\rm eff}$ becomes
significantly weaker, indicating that the intrinsic relation is at most mild.

Finally, the relation between flare amplitude and ${\rm H_\alpha}$ colour excess
is consistent with the well-established link between magnetic activity and
chromospheric emission \citep[e.g.][]{2020ApJ...905..107M}. Stars with larger
(more negative) $E(m_{\rm H\alpha}-m_{\rm CH\alpha})$ values tend to exhibit
higher flare amplitudes, although the correlation remains weak once
observational biases are taken into account.

\subsection{Flare temperature} 

\begin{figure*}[!t]
\begin{center}
\includegraphics[width=160mm]{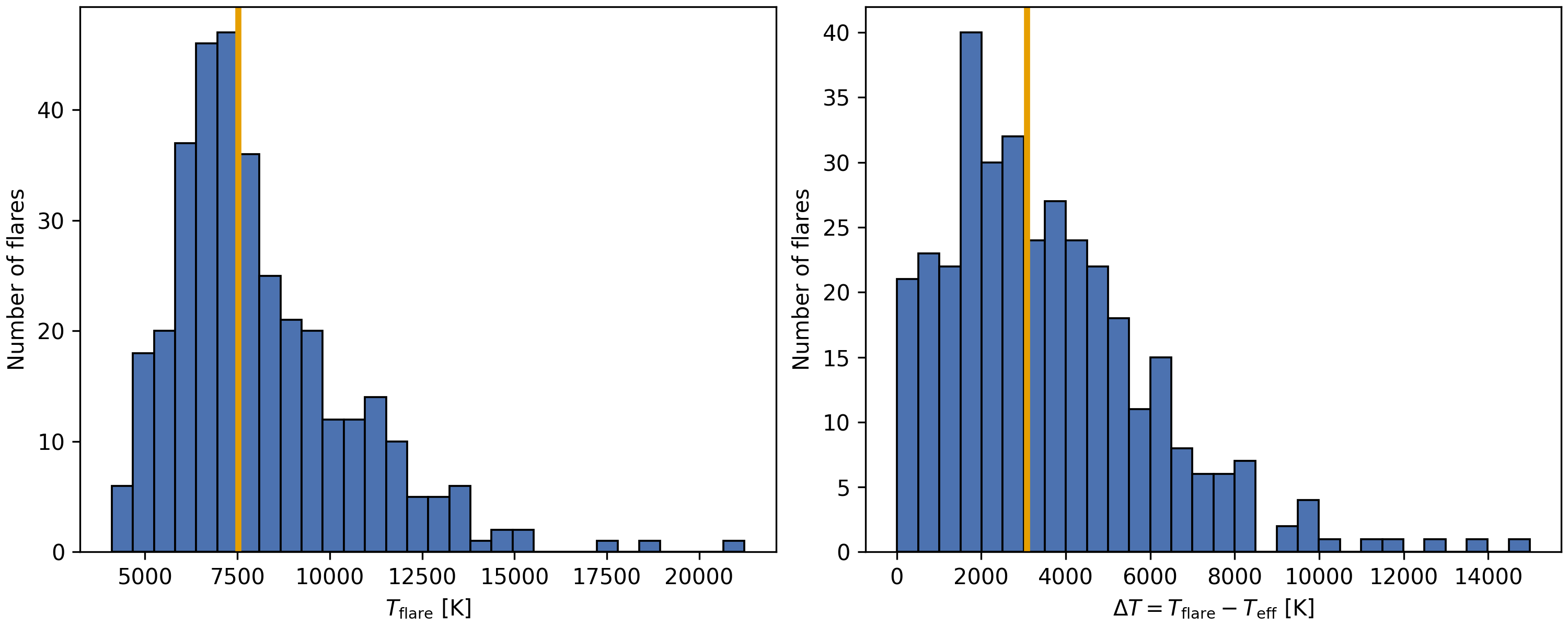}
\caption{Distribution of inferred flare black-body temperatures 
$T_{\rm flare}$ (left panel) and of the temperature contrast between the flare emission and the stellar photosphere, defined as $\Delta T=T_{\rm flare}-T_{\rm eff}$
(right panel). Vertical lines indicate the median values of the distributions.}
\label{histoflare}
\end{center}
\end{figure*}

We computed the flare black-body temperatures following the procedure described in Sect.~\ref{sec:tcomputation}. The left panel of Fig.~\ref{histoflare} shows the distribution of the inferred $T_{\rm flare}$ values. The estimated temperatures span the range from $4\,100$~K to $21\,200$~K, with a peak at $\sim 7\,250$~K and a median value of $\sim 7\,500$~K, both lower than the canonical temperatures of $9\,000$–$10\,000$~K traditionally attributed to stellar flares \citep{1991ApJ...378..725H}. Only about 9\% of the measured values fall within the $9\,000$–$10\,000$~K interval, while approximately 20\% exceed $10\,000$~K.

These results are consistent with \citet{2022AJ....164..223R} and indicate that stellar flares are characterised by a broad range of black-body temperatures rather than by a single canonical value.

It is important to note, however, that these temperatures are derived from single-epoch \gaia\ measurements and therefore do not necessarily correspond to the peak temperature reached during the flare evolution. Nevertheless, the fact that about 20\% of the events already exceed $10\,000$~K at the observed epoch suggests that the peak temperature distribution likely extends to even higher values.  

The right panel of Fig.~\ref{histoflare} shows the distribution of the temperature contrast between the flare emission and the stellar photosphere, defined as $\Delta T = T_{\rm flare} - T_{\rm eff}$. The distribution peaks at $\sim 1\,750$~K and has a median value of $\sim 3\,000$~K, indicating that flare emission is typically hotter than the underlying photosphere by a few thousand kelvin.  

\begin{figure*}
\begin{center}
\includegraphics[width=160mm]{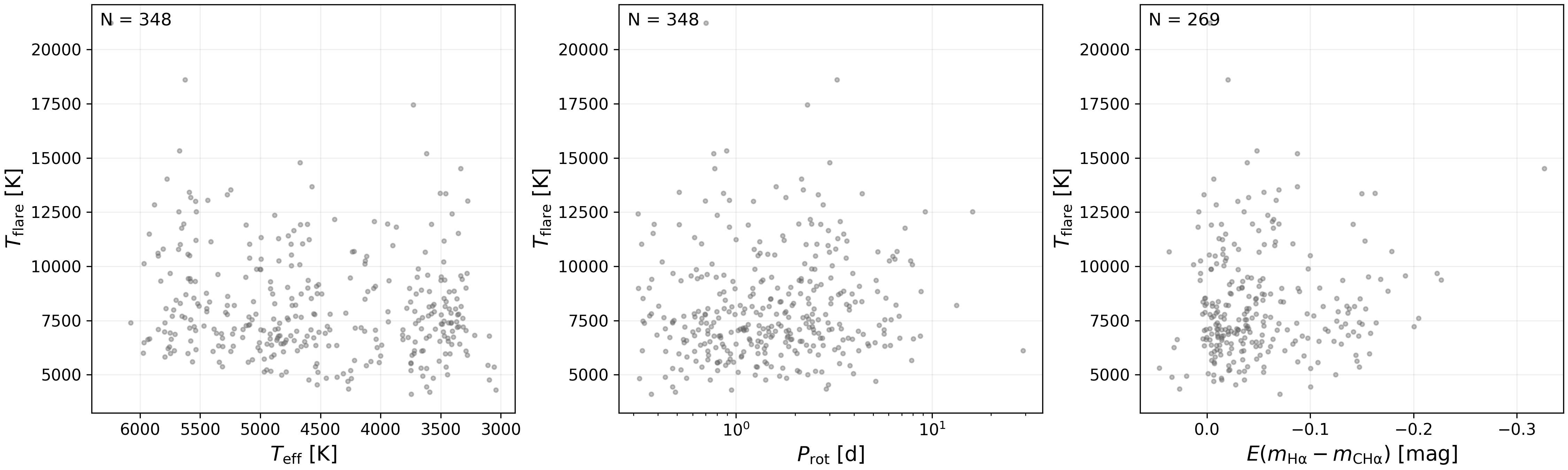}
\caption{Flare effective temperatures $T_{\rm flare}$ as a function of stellar effective temperature (left), rotation period (centre), and $H_{\alpha}$ colour excess (right). Grey points show individual flares. The number of points used in each panel is indicated in the corresponding subplot and reflects the availability of the relevant stellar parameters . No clear trends are apparent in any of the relations.}
\label{tflarepar}
\end{center}
\end{figure*}

 To assess possible dependencies on stellar properties, we examined the relation between $T_{\rm flare}$ and stellar effective temperature, rotation period, and ${\rm H_\alpha}$ colour excess (Fig.~\ref{tflarepar}). A visual inspection reveals no clear systematic variations.
 We quantified these relations by computing the Spearman rank correlation coefficients between $T_{\rm flare}$ and the considered stellar parameters. The resulting values (Table~\ref{spearmancoeff}) confirm the absence of statistically significant correlations, indicating that flare temperature shows no strong dependence on the global stellar parameters explored here.

\begin{figure*}[!t]
\begin{center}
\includegraphics[width=160mm]{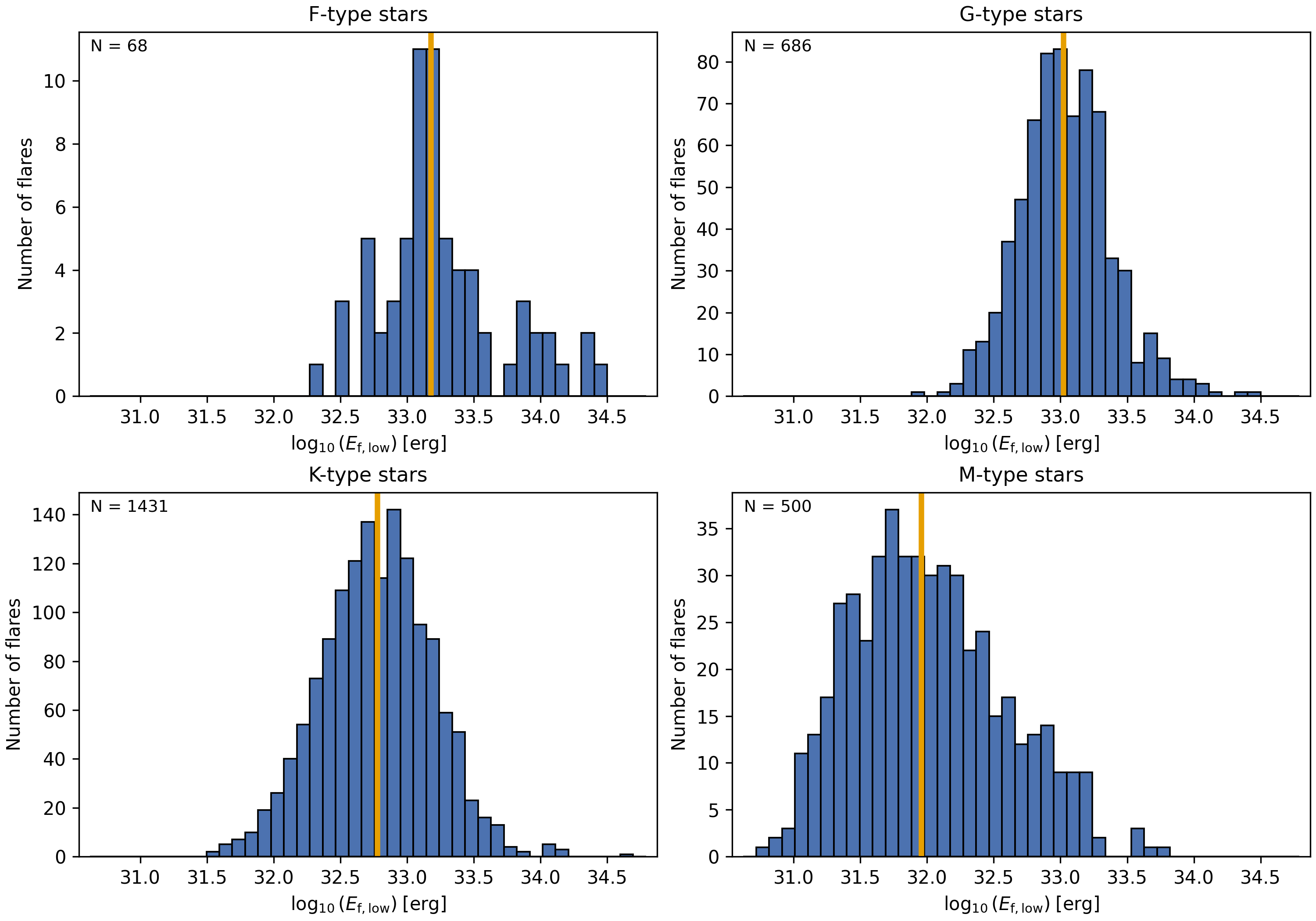}
\caption{$E_{\rm f,low}$ distribution for F-, G-, K-, and M-type stars. 
Spectral types are assigned based on the effective temperatures derived from
the GSP-Phot module, using the calibration of \citet{2013ApJS..208....9P}.
Vertical lines indicate the median values of the distributions. The number of
events in each panel is reported within the corresponding subplot.
The histogram representation is adopted to highlight and compare the energy
distributions across different spectral types.}
\label{histoenergy}
\end{center}
\end{figure*}

\subsection{Flares energies}

We estimated the lower-limit energy $E_{\rm f,low}$ according to
Eq.~\ref{lowerlimit} for 2\,688 flare events.

This corresponds to the subset of sources with available distance estimates
(see Sect.~\ref{sec:energy}). As discussed in Sect.~\ref{sec:energy}, this
quantity represents a lower bound to the energy released during the eruption,
owing to the limited temporal sampling and to the fact that only a fraction of
the flare emission is captured within the \gaia\ photometric bands.

The derived $E_{\rm f,low}$ values span the range $5.8\times10^{30}$ to
$4\times10^{34}$~erg.
To investigate the energy distribution as a function of
spectral type, we assigned spectral types from the effective temperatures
derived by GSP-Phot, using the $T_{\rm eff}$--spectral-type relation tabulated
by \citet{2013ApJS..208....9P}.
Figure~\ref{histoenergy} shows the distribution of $E_{\rm f,low}$ for F-, G-,
K-, and M-type stars.

A shift of the distributions towards higher energies for earlier spectral
types is apparent. However, this behaviour is likely influenced by
observational biases, as flares on more luminous stars must reach higher
energies to be detected above the photometric noise level.

This interpretation is consistent with previous studies based on TESS and K2
data \citep[e.g.][]{2023A&A...669A..15Y,2019ApJ...873...97L}.

We note that a small fraction ($\sim$3--4\%) of the sample may consist of
pre-main-sequence stars; however, this does not significantly affect the
overall distributions.

Among the G-type stars in our sample, 364 flares exhibit $E_{\rm f,low}>10^{33}$~erg. 

Even considering that $E_{\rm f,low}$ provides only a lower limit, these events lie
well above the typical energy range of solar flares ($10^{29}$–$10^{32}$~erg).

Following the definition of \citet{2013ApJS..209....5S}, they can therefore be
classified as candidate super-flares, although their true energies may be
significantly higher than the reported values.

\subsection{Hyper-flares}

Several recent works have focused on \emph{hyper-flares}, which, following
\citet{2018ApJ...867...78C}, are defined as flare events during which the
stellar brightness increases by more than a factor of two. In the \gaia\ $G$
band, this corresponds to amplitudes $A(G)\ge 0.75\,\mathrm{mag}$. Such
high-amplitude events have been reported to occur preferentially in M dwarfs,
whose deep convective envelopes sustain efficient turbulent dynamos capable of
producing powerful magnetic flares \citep{2014ApJ...797..122D}.

In the present work, we detect 29 hyper-flares occurring on 28 stars (two
events are detected for a single source). Stellar effective temperatures are
available for 26 of these stars: 21 are M dwarfs with
$T_{\rm eff}<4\,000$~K, while five objects are
late-K dwarfs with $T_{\rm eff}$ in the range $\sim$4\,050--4\,650~K.
Hyper-flares are highlighted in Fig.~\ref{ampvs.par} by the horizontal dashed
line corresponding to the adopted amplitude threshold.

Although these events lie outside the completeness-limited regime defined in
Sect.~4.3, this does not affect their detectability, as completeness primarily
impacts low-amplitude flares.

Visual inspection of Fig.~\ref{ampvs.par} indicates that hyper-flares
preferentially occur in cool stars and are often associated with strong
${\rm H_\alpha}$ emission, consistent with previous findings
\citep[e.g.][]{2018ApJ...867...78C}. For 13 hyper-flare host stars we were able
to compute the photometric index $E(m_{\rm H\alpha}-m_{\rm CH\alpha})$, and ten
of these show strong ${\rm H_\alpha}$ emission, with
$E(m_{\rm H\alpha}-m_{\rm CH\alpha})<-0.12$.

The limited number of hyper-flare host stars prevents a
statistically meaningful comparison of their properties with those of the
overall population of flaring stars. Nevertheless, the predominance of M
dwarfs is consistent with the expected increase in magnetic activity towards
lower stellar masses. The presence of a few late-K stars is not unexpected, as
stars in this temperature range still possess convective envelopes capable of
sustaining strong magnetic activity, although such extreme events are expected
to be less frequent than in M dwarfs.

Inspection of their position in the HR diagram suggests that one of these
late-K objects is compatible with a pre-main-sequence classification, one lies
close to the binary sequence, while the remaining three are consistent with
main-sequence stars.

Finally, for three hyper-flare events we were able to estimate the flare
effective temperature $T_{\rm flare}$. Two occurred on M2 stars with inferred
temperatures of $7\,767$~K and $7\,448$~K, respectively, while the third event,
detected on a K6 star, has an estimated temperature of $10\,111$~K.

A more detailed investigation of these rare events is beyond
the scope of the present work and would require dedicated follow-up
observations.

\section{Conclusions} 

In this work we presented a method for the detection and characterisation of
stellar flares in \gaia\ photometric time series, exploiting both the
multi-band photometry and the low-resolution XP spectra. In particular, we
introduced a dedicated flare-detection pipeline and defined the photometric
index $E(m_{\rm H\alpha}-m_{\rm CH\alpha})$ to characterise ${\rm H_\alpha}$
emission using \gaia\ spectrophotometric data.

Applying this approach to the {\tt gdr3\_rotmod} catalogue, we detected 3\,217
flares occurring on 2\,818 stars out of approximately 474\,000 analysed
sources. The relatively small fraction of stars showing detectable flares is
consistent with the intrinsically intermittent nature of flaring activity and
with the sparse temporal sampling of \gaia. Nevertheless, this result
demonstrates that \gaia\ can identify a statistically significant population
of stellar flares in an all-sky, homogeneous stellar sample.

A careful assessment of detection completeness allowed us to identify a
magnitude range in which the flare detection sensitivity, and therefore the
minimum detectable amplitude, is approximately uniform. We then investigated
the dependence of the $G$-band flare amplitude, $A(G)$, on stellar properties
by computing Spearman rank correlation coefficients for both the full sample
and the completeness-limited sample.

In the full sample, we find mild correlations between $A(G)$ and both stellar
effective temperature and ${\rm H_\alpha}$ colour excess, with typical values
$|r_s|\simeq 0.4$. However, these correlations are significantly reduced when
restricting the analysis to the completeness-limited sample, where
$|r_s|\simeq 0.2$. This demonstrates that the stronger correlations observed
in the full sample are largely driven by magnitude-dependent observational
biases. In the completeness-limited sample, the dependence of flare amplitude
on stellar parameters is weak and accompanied by a large intrinsic dispersion.

The correlation between flare amplitude and rotation period is weak in both
the full and completeness-limited samples. This is primarily due to the strong
bias of the \gaia\ rotational-modulation catalogue towards fast rotators, with
approximately 90\% of the flaring stars having $P_{\rm rot}\le 5\,\mathrm{d}$.
As a consequence, the limited range of rotation periods
covered by our sample may reduce our sensitivity to a possible dependence of
flare amplitude on rotation period.

The multi-band \gaia\ photometry further enabled us to estimate effective flare
temperatures for a subset of 348 events using a black-body approximation.
The distribution of flare temperatures peaks around
$7\,500~\mathrm{K}$, supporting recent results that challenge the assumption
of a single characteristic flare temperature in the
$9\,000$--$10\,000~\mathrm{K}$ range
\citep{2022AJ....164..223R}.
These estimates are based on single-epoch observations and therefore do not
necessarily correspond to the peak temperature reached during the flare
evolution. However, the presence of events with inferred temperatures well
above $10\,000~\mathrm{K}$ already at the observed epoch indicates that the
peak temperature distribution likely extends to significantly higher values.
We also examined the temperature contrast between the flare emission and the
stellar photosphere, finding typical contrasts of a few thousand kelvin. No
statistically significant correlations between $T_{\rm flare}$ and stellar
effective temperature, rotation period, or ${\rm H_\alpha}$ emission are
found.

Finally, the distribution of flaring stars in the Hertzsprung--Russell diagram
shows that they are preferentially located above the main sequence,
whereas the {\tt gdr3\_rotmod} parent sample is distributed
on both sides of it. This difference is particularly interesting in light of
the results of \citet{2023AJ....166...63J}, who found that, for M dwarfs,
fast rotators and H$\alpha$-active stars show similar distributions in the HR
diagram, with both populations preferentially located above the main sequence.
The distribution observed for our flaring stars may therefore indicate that
they trace the more magnetically active component of the
{\tt gdr3\_rotmod} population, although the physical origin of their
displacement above the main sequence is still not fully understood.

Owing to the limited temporal sampling of \gaia, flare durations and total
emitted energies cannot be directly measured. We therefore estimated a strict
lower limit to the flare energy, $E_{\rm f,low}$. The inferred energy
distributions show a systematic shift towards higher values for earlier
spectral types; however, this behaviour is likely influenced by detection
biases, as flares on more luminous stars must reach higher energies to be
detected. In particular, we identify 364 events on G-type stars with
$E_{\rm f,low} > 10^{33}\,\mathrm{erg}$, which can be classified as
super-flares. We also detected 29 hyper-flares, defined as
events with amplitudes $A(G)\ge 0.75\,\mathrm{mag}$, which preferentially occur
in cool stars and are frequently associated with strong ${\rm H_\alpha}$
emission.

A key strength of \gaia\ lies in the discovery of flaring stars across the
entire sky. Among the flaring sources identified in this work, only 14 were
previously reported in the literature, while the vast majority are classified
here as flaring variables for the first time. This highlights the unique
capability of \gaia\ to uncover new flaring stars despite its sparse temporal
sampling.

Although \gaia\ is not optimised for time-resolved flare studies, its all-sky
coverage combined with simultaneous multi-band photometry and
spectrophotometric information provides a valuable and complementary
perspective to high-cadence missions such as {\it Kepler} and TESS, enabling
the characterisation of flare properties for a subset of events.

In future data releases, the algorithm described here will be applied to
broader stellar samples, extending the present analysis and enabling a more
complete and homogeneous all-sky census of stellar flaring activity across
different stellar populations.

\begin{acknowledgements}
\addcontentsline{toc}{chapter}{Acknowledgements}

This work presents results from the European Space Agency (ESA) space mission \gaia. \gaia\ data are being processed by the Institutions participating in the \gaia\ MultiLateral Agreement (MLA). The \gaia\ mission website is \url{https://www.cosmos.esa.int/gaia}. The \gaia\ archive website is \url{https://archives.esac.esa.int/gaia}.
The \gaia\ mission and data processing have financially been supported by: the Agenzia Spaziale Italiana (ASI) through contracts I/037/08/0, I/058/10/0, 2014-025-R.0, 2014-025-R.1.2015, and 2018-24-HH.0 to the Italian Istituto Nazionale di Astrofisica (INAF), contract 2014-049-R.0/1/2 to INAF for the Space Science Data Centre (SSDC, formerly known as the ASI Science Data Center, ASDC), contracts I/008/10/0, 2013/030/I.0, 2013-030-I.0.1-2015, and 2016-17-I.0 to the Aerospace Logistics Technology Engineering Company (ALTEC S.p.A.), INAF, and the Italian Ministry of Education, University, and Research (Ministero dell'Istruzione, dell'Universit\`{a} e della Ricerca) through the Premiale project `MIning The Cosmos Big Data and Innovative Italian Technology for Frontier Astrophysics and Cosmology' (MITiC); the Swiss State Secretariat for Education, Research and Innovation through the ``Activit\'{e}s Nationales Compl\'{e}mentaires’'.

The \gaia\ project and data processing have made use of: the Set of Identifications, Measurements, and Bibliography for Astronomical Data \citep[SIMBAD,][]{2000A&AS..143....9W}, the `Aladin sky atlas' \citep{2000A&AS..143...33B,2014ASPC..485..277B}, and the VizieR catalogue access tool \citep{2000A&AS..143...23O}, all operated at the Centre de Donn\'{e}es astronomiques de Strasbourg (\href{http://cds.u-strasbg.fr/}{CDS}); the software products \href{http://www.starlink.ac.uk/topcat/}{TOPCAT}, \href{http://www.starlink.ac.uk/stil}{STIL}, and \href{http://www.starlink.ac.uk/stilts}{STILTS} \citep{2005ASPC..347...29T,2006ASPC..351..666T}; Matplotlib \citep{Hunter:2007}; Astropy, a community-developed core Python package for Astronomy \citep{2018AJ....156..123A};
\label{lastpage}
\end{acknowledgements}

\bibliographystyle{aa}
\bibliography{SDRref}
\begin{appendix}
\section{Computation of the  $E(m_{\rm H\alpha}-m_{\rm CH\alpha})$ index}
\label{expectedhalpha}
\subsection{Computation of the expected $m_{H\alpha}-m_{CH\alpha}$ colour. }
\begin{figure}

\begin{center}
\includegraphics[width=80mm]{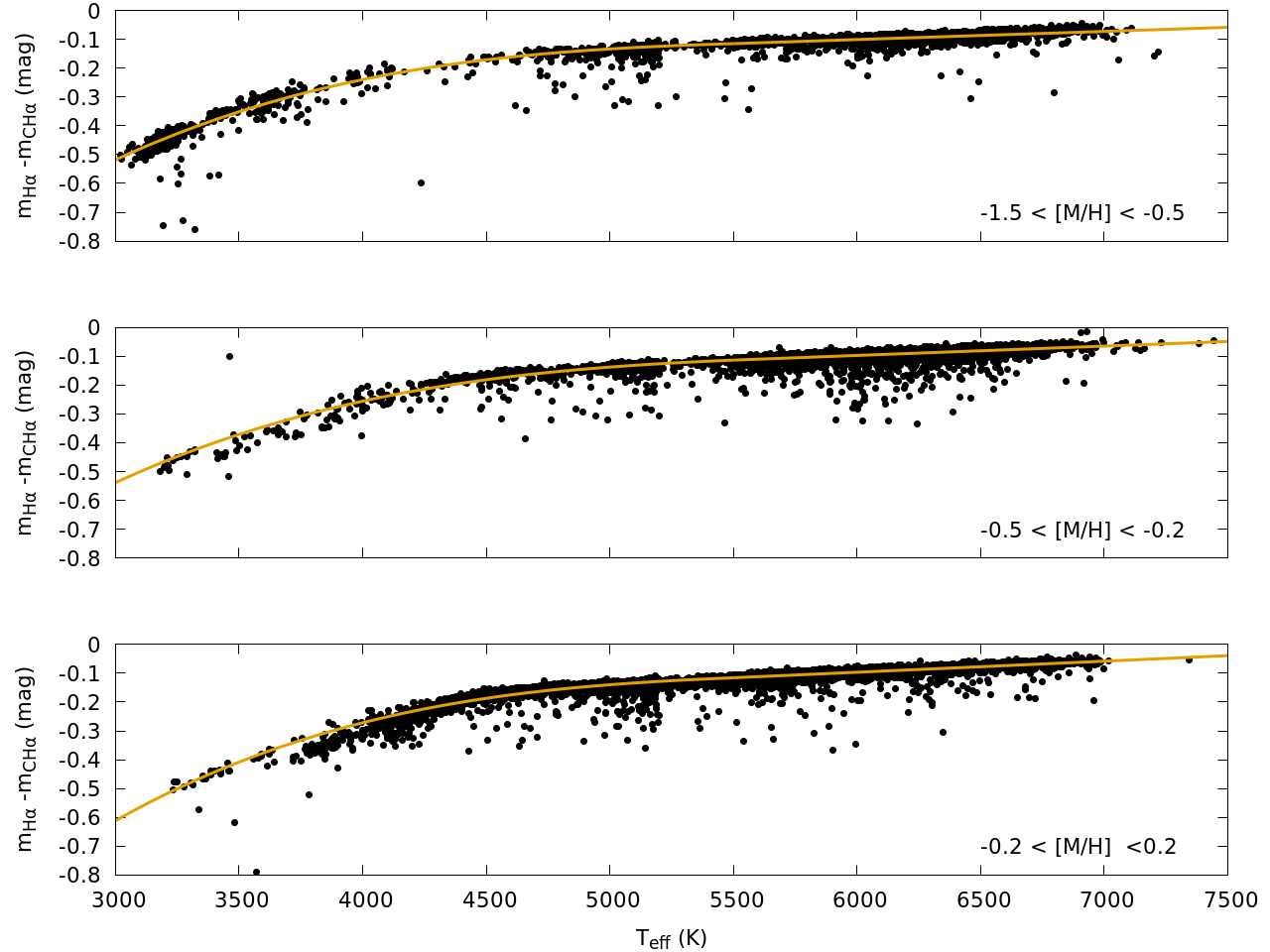}
\caption{Observed $(m_{H\alpha}-m_{CH_\alpha})$ colour vs. $T_{\rm eff}$ for the sample of magnetically inactive stars taken from \cite{2021A&A...649A...2L}. Top: stars with -1.5 < [M/H] <-0.5. Middle: stars with -0.5<[M/H]<-0.2. Bottom: -0.2< [M/H] <0.2. }
\label{polifit}
\end{center}
\end{figure}

We selected a sample of magnetically inactive stars from the catalogue compiled by \cite{2023A&A...674A..30L} and stored in the {\tt gaiadr3.astrophysical\_parameters} table  of the \text{Gaia} main database. This table reports the magnetic activity index $\alpha$  for about two million sources. Such an index, inferred by the analysis of the Ca II infrared triplet (IRT) spectral lines observed by the \text{Gaia} Radial Velocity Spectrometer (RVS), is a measure of the excess equivalent width factor in
the core of the Ca II IRT lines with respect to a reference inactive spectrum. The closer $\alpha$ is to 0, the lower the stellar magnetic activity.
We selected the sample of inactive stars by employing the following criteria:

\begin{equation}
 \left\{
\begin{array}{l}
	|\alpha \pm 3\sigma_\alpha| < 0.003 	\\\\
		      
        \mbox{if }  T_{\rm eff} \ge 5\,000 ~\rm K \\\\
        
      |\alpha \pm 3\sigma_\alpha|< 0.009   \\
		 \\ \mbox{if } T_{\rm eff} < 5\,000~ \rm K
\end{array}
\right .
\label{crithalpha}
\end{equation}

where $\alpha$, $\sigma_\alpha$ and $T_{\rm eff}$ are the values of the parameters {\tt activityindex\_espcs}, {\tt activityindex\_espcs\_uncertainty} and {\tt teff\_gspphot} stored  in the table {\tt gaiadr3.astrophysical\_parameters}, respectively.
The different thresholds used to select the inactive stars are due to the fact that $\alpha$ also depends on the spectral type and tends to have higher values towards cooler stars.  \citep[See][for further details] {2023A&A...674A..30L}.
Two further criteria were added to restrict the sample of inactive stars to the gravity and metallicity range in which lie our flaring stars:
\begin{equation}
    \label{critmetal}
    -1.53 \le [M/H] \le 0.23
\end{equation}
\begin{equation}
\label{critlogg}
3.98\le {\rm log}~g \le 4.89
\end{equation}

where [{\rm M/H}] and $ {\rm log} g$ are the {\tt mh\_gspphot} and {\tt logg\_gspphot} parameters listed in the {\tt gaiadr3.astrophysical\_parameters}.
The upper and lower thresholds used in the above criteria are the $1^{st}$ and $99^{th}$ percentile of the [M/H] and $ {\rm log} g$  distributions of our flaring stars.
We employed Gaia-XPy to extract the $m_{H\alpha} -m_{CH\alpha}$ colour for the inactive stars selected with the above criteria.
In Fig. \ref{polifit} we plotted $m_{H\alpha} -m_{CH\alpha}$ against $T_{\rm eff}$ for three different ranges of metallicity.
We fitted the upper envelope of the three datasets with the function:

\begin{equation}
(m_{H\alpha} -m_{CH\alpha}) =
 \left\{
\begin{array}{l}
		c_0 +c_1 T_{\rm eff} +c_2 T_{\rm eff}^2 +c_3 T_{\rm eff}^3 \\\\
      
        \mbox{if } 3\,000 ~\rm K <T_{\rm eff} <5\,500 ~\rm K \\
        \\
		b_0 +b_1 T_{\rm eff} \\\\ \mbox{if } 5\,500 ~\rm K <T_{\rm eff} <7\,500~\rm K
\end{array}
\right .
\label{fitqeuation}
\end{equation}

The fitted curves are over-plotted on the data.
The coefficients of the fitted polynomials are reported in Tables \ref{polifitcoeff} and \ref{polifitseg} together with the metallicity and the temperature range in which they are valid.
These empirical relations can be used to infer the expected $m_{H\alpha}-m_{CH\alpha}$ colour for a star of known $T_{\rm eff}$ and metallicity [M/H]. 

\begin{table}
\caption{Fit parameters in the $T_{\rm eff}$ range $(5500:7500 ~\rm K)$ }
\centering
\begin{tabular}{lllll}
\hline
[M/H] range  & $c_0$ &  $c_1$ &$c_2$ &$c_3$  \\
\hline 
(-1.5:-0.5)   & -3.13 & 0.0015& -2.5e-7 & 1.4e-11     \\
(-0.5:-0.2)   & -2.92 &0.0013&-2.06e-7&1.1e-11     \\
(-0.2:0.2)   & -3.12 &0.0014& -2.12e-7&1.1e-11    \\
\hline
\end{tabular}
\label{polifitcoeff}
\end{table}

\begin{table}
\caption{Fit parameters in the $T_{\rm eff}$ range $(3000:5500 ~\rm K)$ }
\centering
\begin{tabular}{lll}
\hline
[M/H] range  & $b_0$ &  $b_1$ \\
\hline 
(-1.5:-0.5)   & -0.27 &  2.82e-5 \\

(-0.5:-0.2)   & -0.29 &3.23e-5\\
(-0.2:0.2)   & -0.33 &3.83e-5\\
\hline
\end{tabular}
\label{polifitseg}
\end{table}

\section{Comparison with previous works}
\label{sec:previousworks}

\begin{figure}
\begin{center}
\includegraphics[width=80mm]{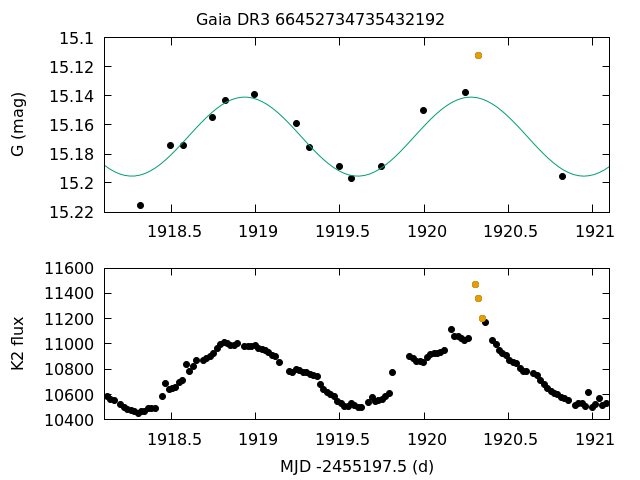}
\caption{Over-lapping \gaia and K2  data-sets for the star {\tt Gaia DR3 66452734735432192}. The yellow points mark the flare detected in both time-series. Top panel: $G$ mag measurements. The harmonic-model best-fitting the data is over-plotted on the data (green line).  Bottom panel: K2 flux measurements.   }
\label{k2gaia_ex2}

\end{center}
\end{figure}

\begin{figure}
\begin{center}
\includegraphics[width=80mm]{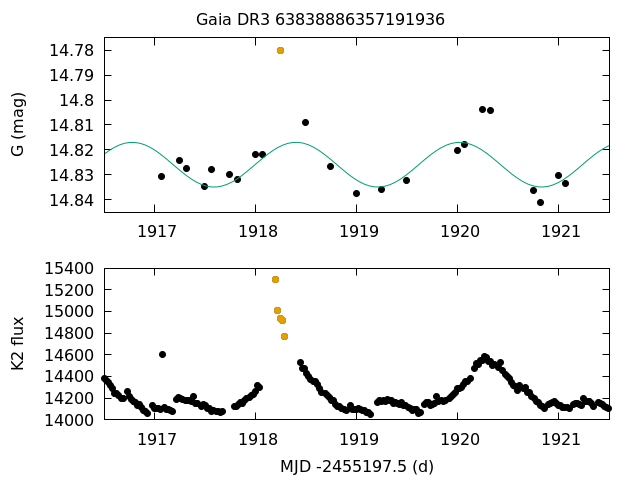}
\caption{Same as Fig. \ref{k2gaia_ex2} for the star {\tt Gaia DR3 63838886357191936}.}
\label{k2gaia_ex1}

\end{center}
\end{figure}

\begin{figure}
\begin{center}
\includegraphics[width=80mm]{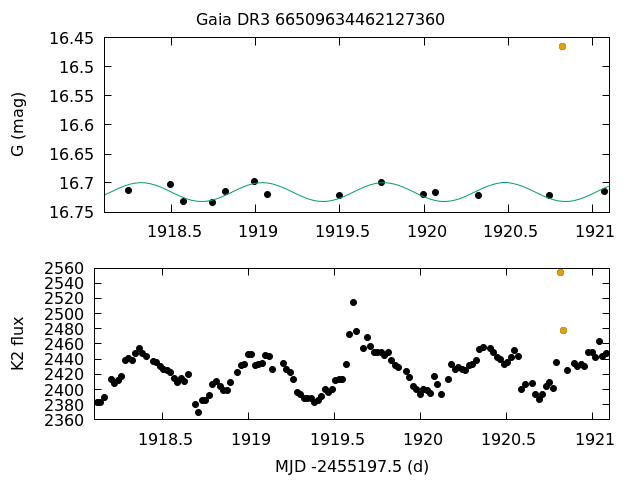}
\caption{Same as Fig. \ref{k2gaia_ex2} for the star {\tt Gaia DR3 66509634462127360}. Note that in this case the flare, though visible in K2 data, was detected only in the present work. }
\label{k2gaia_ex3}

\end{center}
\end{figure}

We cross-matched the catalogue of the flaring stars detected here with the SIMBAD  astronomical database \citep{2000A&AS..143....9W} at CDS. We found that 14 stars out of our sample have been classified as flaring stars in previous papers. These stars have been listed in Table \ref{previousworks}  together with the references to the previous works. 
Note that 7 stars out of those reported in Table \ref{previousworks} are members of the Pleiades cluster surveyed by the campaign 4 of the K2  mission and studied by \cite{2021A&A...645A..42I}. For 3 of these stars, the \gaia time-series partially over-lap with the K2 time-series analysed by \cite{2021A&A...645A..42I} and the flares detected in \gaia data are clearly visible in the  30 minutes cadence K2 time-series.  We reported these over-lapping datasets in Fig. \ref{k2gaia_ex2}, \ref{k2gaia_ex1} and \ref{k2gaia_ex3} for the stars  {\tt Gaia DR3 66452734735432192},{\tt Gaia DR3 6383886357191936}, and   {\tt Gaia DR3 66509634462127360}, respectively. 
In each picture we reported the \gaia $G$ time-series (top panel)  and the corresponding K2 time-series (bottom panel).  The harmonic model best-fitting the \gaia data is over-plotted on the $G$ time-series. The yellow points are used to mark the flare event detected in both data-sets. Despite that the \gaia scanning-law does not permit to sample the temporal evolution of the flare, our algorithm is able to detect it.

Note that the flares marked in Fig. \ref{k2gaia_ex2} and Fig. \ref{k2gaia_ex1} were both detected by \cite{2021A&A...645A..42I}, whereas the flare marked in Fig. \ref{k2gaia_ex3} was not reported by the same authors. Nevertheless the flare is clearly visible in K2 time-series and it was not reported by \cite{2021A&A...645A..42I} because their detection algorithm searches only  events in which at least three consecutive data points lie $3\sigma$ above the quiet flux.

\begin{table}[!t]
\caption{Flaring stars reported in previous studies}
\label{previousworks}
\centering
\scriptsize
\setlength{\tabcolsep}{2pt}
\begin{tabular}{@{}r l l@{}}
\hline\hline
\gaia DR3 sourceid & Other designation & Ref.\\
\hline
51526417710225408  & EPIC 210840112 & \text{\cite{2021A&A...645A..42I}}\\
63838886357191936  & EPIC 210939348 & \text{\cite{2021A&A...645A..42I}}\\
63892006512622848  & EPIC 210971348 & \text{\cite{2021A&A...645A..42I}}\\
63959351598814336  & EPIC 210949721 & \text{\cite{2021A&A...645A..42I}}\\
66452734735432192  & EPIC 211046168 & \text{\cite{2021A&A...645A..42I}}\\
66509634462127360  & EPIC 211059754 & \text{\cite{2021A&A...645A..42I}}\\
159559619890954752 & HAT 216-04245  & \text{\cite{2011AJ....141..166H}}\\
661402015575639680 & EPIC 211982734 & \text{\cite{2021A&A...645A..42I}}\\
1851868127125091840 & Ross 776      & \text{\cite{2020ApJ...892..144R}}\\
1923896725839633024 & G 190-28      & \text{\cite{2014AcA....64..359T}}\\
5088895681454956672 & TIC 121011020 & \text{\cite{2020ApJ...890...46T}}\\
5382868669400908928 & V* V857 Cen   & \text{\cite{1999A&AS..139..555G}}\\
6504144270854070272 & TIC 231708319 & \text{\cite{2020AJ....159...60G}}\\
6824372698820338176 & TIC 25243422  & \text{\cite{2020AJ....159...60G}}\\
\hline
\end{tabular}
\end{table}
\FloatBarrier

\end{appendix}

\end{document}